\documentclass[a4paper, 11pt]{article}
\usepackage[top = 2cm, bottom = 2cm, left = 2cm, right = 2cm]{geometry} 

\usepackage[default]{cantarell}
\usepackage{setspace}
\usepackage[utf8]{inputenc}

\usepackage{graphicx} 

\usepackage{hyperref}
\usepackage{empheq}
\usepackage{ulem} 

\usepackage{dsfont}
\usepackage{amsmath}
\usepackage{amsfonts}

\usepackage{colortbl}
\usepackage[table]{xcolor}
\definecolor{myblue}{RGB}{17,85,204}

\newcommand{\bfs}[1]{\boldsymbol{#1}}

\newcommand{\RVW}[1]{\textcolor{black}{#1}}

\begin{document}

\begin{center}
    \large \bf Structuring, Sequencing, Staging, Selecting: the 4S method for the longitudinal \\ 
    analysis of multidimensional questionnaires in chronic diseases
\end{center}

\begin{center}
    {\bf Tiphaine Saulnier$^{1,*}$, Wassilios G. Meissner$^{2,3,4}$, Margherita Fabbri$^{5,6}$, \\ Alexandra Foubert-Samier$^\mathbf{\dagger1,2,3,7}$, and Cécile Proust-Lima$^\mathbf{\dagger1,7}$}
\end{center}

\vspace{.25cm}
\small
\noindent$^1$Univ. Bordeaux, Bordeaux Population Health research center, Inserm U1219, Bordeaux, France \\
$^2$CHU Bordeaux, Service de Neurologie des Maladies Neurodégénératives, CRMR AMS, NS-Park/FCRIN Network, Bordeaux, France \\
$^3$Univ. Bordeaux, IMN, CNRS, UMR 5293, Bordeaux, France \\
$^4$Dept. Medicine, Univ. Otago, New Zealand Brain Research Institute, Christchurch, New Zealand \\
$^5$CHU Toulouse, MSA French Reference Center, Toulouse, France \\
$^6$Univ. Toulouse, Dept. Clinical Pharmacology and Neurosciences, CIC-1436, NeuroToul COEN Center, NS-Park/FCRIN Network, Inserm U1048/1214, Toulouse, France \\
$^7$Inserm, CIC1401-EC, Bordeaux, France \\
$^\dagger$These authors contributed equally. \\
*\textit{email: tiphaine.saulnier@u-bordeaux.fr}

\normalsize
\vspace{1cm}


\noindent \textbf{abstract} \\
    In clinical studies, questionnaires are often used to report disease-related manifestations from clinician and/or patient perspectives. Their analysis can help identify relevant manifestations throughout the disease course, enhancing knowledge of disease progression and guiding clinicians in appropriate care provision.
    However, the analysis of questionnaires in health studies is not straightforward as made of repeated, ordinal, and potentially multidimensional item data. Sum-score summaries may considerably reduce information and hamper interpretation; items' changes over time occur along clinical progression; and as many other longitudinal processes, observations may be truncated by events.
    This work establishes a comprehensive strategy in four consecutive steps to leverage repeated ordinal data from multidimensional questionnaires. The 4S method successively (1) identifies the questionnaire structure into dimensions satisfying three calibration assumptions (unidimensionality, conditional independence, increasing monotonicity), (2) describes each dimension progression using a joint latent process model which includes a continuous-time item response theory model for the longitudinal subpart, (3) aligns each dimension progression with disease stages through a projection approach, and (4) identifies the most informative items across disease stages using the Fisher information.
    The method is applied to multiple system atrophy (MSA), a rare neurodegenerative disease, with the analysis of daily activity and motor impairments over disease progression. The 4S method provides an effective and complete analytical strategy for questionnaires repeatedly collected in health studies.

\vspace{.5cm}
\noindent \textbf{keywords} \\
Item response theory; Joint modeling; Multidimensional questionnaires; Multiple system atrophy; Multivariate longitudinal data; Neurodegenerative diseases.

\newpage

\section{Introduction} \label{s:intro}

\RVW{The study of chronic diseases increasingly involves the longitudinal analysis of Clinical Outcome Assessments (COAs), which enumerate disease-related manifestations through questionnaires. These include patient-reported outcomes (PRO), clinician-reported outcomes (ClinRO), observer-reported outcomes (ObsRO) or performance outcomes (PerfO). Their longitudinal analysis can help identify relevant manifestations throughout the disease course, enhance knowledge of disease progression, and guide clinicians in delivering appropriate care. They are also increasingly used as primary outcome in clinical trials \cite{cai_longitudinal_2021}.} \\
COAs are central in neurodegenerative diseases with the assessment of diverse clinical manifestations such as cognitive, behavioral, or functional impairments \cite{mestre_patient-centered_2024}.
\RVW{In multiple system atrophy (MSA), a rare alpha-synucleopathy which motivated this work, disease-related impairments are reported by the Unified MSA Rating Scale (UMSARS), a four-part COA survey \cite{wenning_development_2004}. UMSARS parts I and II, the main endpoints in MSA clinical trials, evaluate the disease impact on daily activities and on the motor function, respectively. They are made of 12 and 14 Likert-type items, each one rating a specific activity or symptom in 5 levels from no impairment to extreme impairment.} \\

The longitudinal analysis of COAs like UMSARS-I and -II combines statistical challenges due to both the nature of COAs and their repeated assessment over time. First, questionnaires are essentially made of items, most often quoted on a Likert (i.e., ordinal) scale as in MSA, with each item measuring a sensible aspect of the construct of interest \cite{joshi_likert_2015}. Most studies sum them to define sum-scores, including in MSA \RVW{with UMSARS-I and -II total score} \cite{foubert-samier_disease_2020}, although this considerably reduces the information by not capturing item level differences, raises issues with missing items, and relies on often overlooked assumptions such as unequal interval scaling \cite{reeve_psychometric_2007}. The Item Response Theory (IRT) can leverage the full information of a set of ordinal items and retrieve the common underlying construct (or latent trait) of real interest \cite{gorter_why_2015} using equations of observations to link the latent trait to each item \cite{baker_item_2004}. \RVW{Classical IRT relies} though on  assumptions to be carefully checked: all items should measure the same common underlying construct, should be non-redundant, and should provide significant information on the construct of interest \cite{reeve_psychometric_2007}. As aimed to capture different aspects of a disease, questionnaires are mostly multidimensional, and the identification of homogeneous dimensions thus constitutes a prerequisite \cite{bartolucci_dimensionality_2012}. \\
In longitudinal studies where assessments are repeated over time, the within-patient correlation needs to be accounted for to correctly model the latent trait and assess its determinants. In addition, the data collection may be interrupted by \RVW{intercurrent events} (e.g., dropout, death) \cite{lawrance_what_2020}, that induce missing data, usually not at random as more likely among most affected patients. The IRT methodology has been recently extended to longitudinal data by combining a structural linear mixed model at the latent trait level \cite{proust-lima_modeling_2022} and by simultaneously assessing the risk of event according to the underlying construct trajectory in a joint modeling framework \cite{saulnier_joint_2022}. \\
With its extension to longitudinal data, the IRT methodology offers a powerful tool to analyze in depth unidimensional constructs from series of items in chronic diseases. Yet their interpretability remains often too limited and complex for their use in clinical research. Each latent construct is defined according to its own continuum independently from the others, which hampers the understanding of how they deteriorate along disease progression. The projection of a reference progression marker such as disease stages in MSA along the continuums could facilitate the establishment of benchmarks for better contextualization and interpretation. Second, identifying the key items or manifestations driving each construct progression could prove valuable to guide clinicians for the monitoring of the disease, and to develop reduced scales for clinical trials \cite{edelen_applying_2007,saulnier_patient-perceived_2024}. \\

\begin{figure}[h!]
    \caption{\textbf{4S method mnemonic scheme} with \textbf{Step 1 - Structuring}, the identification of the subdimensions of the questionnaire (Section \ref{ss:4S1}), \textbf{Step 2 - Sequencing}, the description of item impairment sequence and associated factors (Section \ref{ss:4S2}), \textbf{Step 3 - Staging}, the contrast of item impairment hierarchy with disease stages (Section \ref{ss:4S3}), and \textbf{Step 4 - Selecting}, the selection of the most informative items by disease stages (Section \ref{ss:4S4}).  The scheme illustrates the 4S method for a fictive questionnaire composed of 8 items with 2 underlying dimensions.}
    \centering
    \includegraphics[width = 17cm, trim= 1cm 0cm 0cm 0cm, clip]{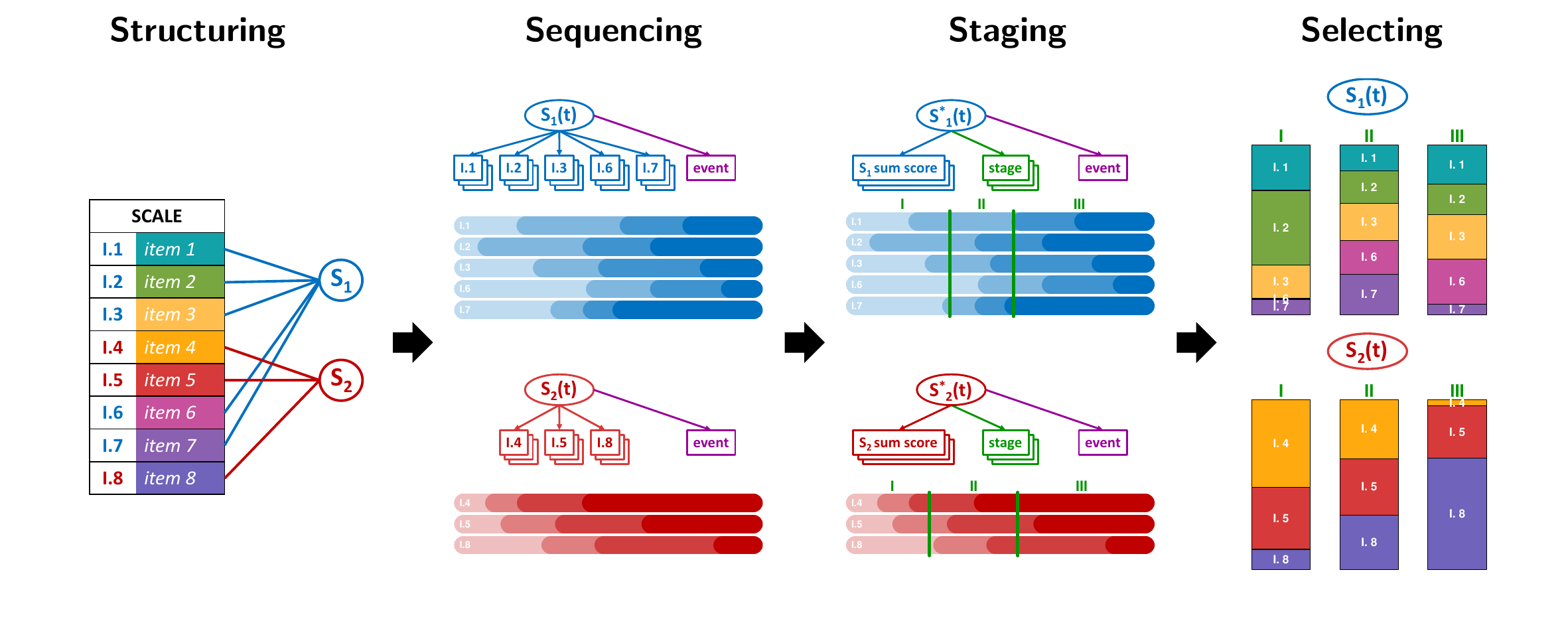}
    \label{fig:4Smethod}
\end{figure}

Despite the growing importance of COAs in health studies, \RVW{there is no comprehensive strategy to analyze longitudinal data of surveys made of Likert items while meeting the statistical challenges they induce.} 
To fill this gap, we propose a four-step strategy, the {\bf 4S method}, schematized in Figure \ref{fig:4Smethod}. It (1) identifies the \textbf{Structure} in dimensions, (2) establishes the \textbf{Sequences} of item impairments and assesses disparities among patients, (3) aligns the dimensions to disease \textbf{Stages}, and (4) \textbf{Selects} the most informative items. \\

Section \ref{s:cohort} presents the French MSA cohort which analysis motivated the 4S method development. Section \ref{s:4S} describes the 4S methodology step-by-step. Section \ref{s:appli} illustrates its application to describe the progression of MSA using the 26 items of UMSARS-I and -II. Finally, Section \ref{s:discuss} concludes. 


\section{The French MSA cohort (FMSA)} \label{s:cohort}

\begin{figure}
    \caption{\textbf{Structuration of UMSARS-I and -II items into dimensions across 50 random replicates (n=731).}}
    \centering
    \includegraphics[width = 17cm, trim= 0cm 2.5cm 7cm 0cm, clip, page=1]{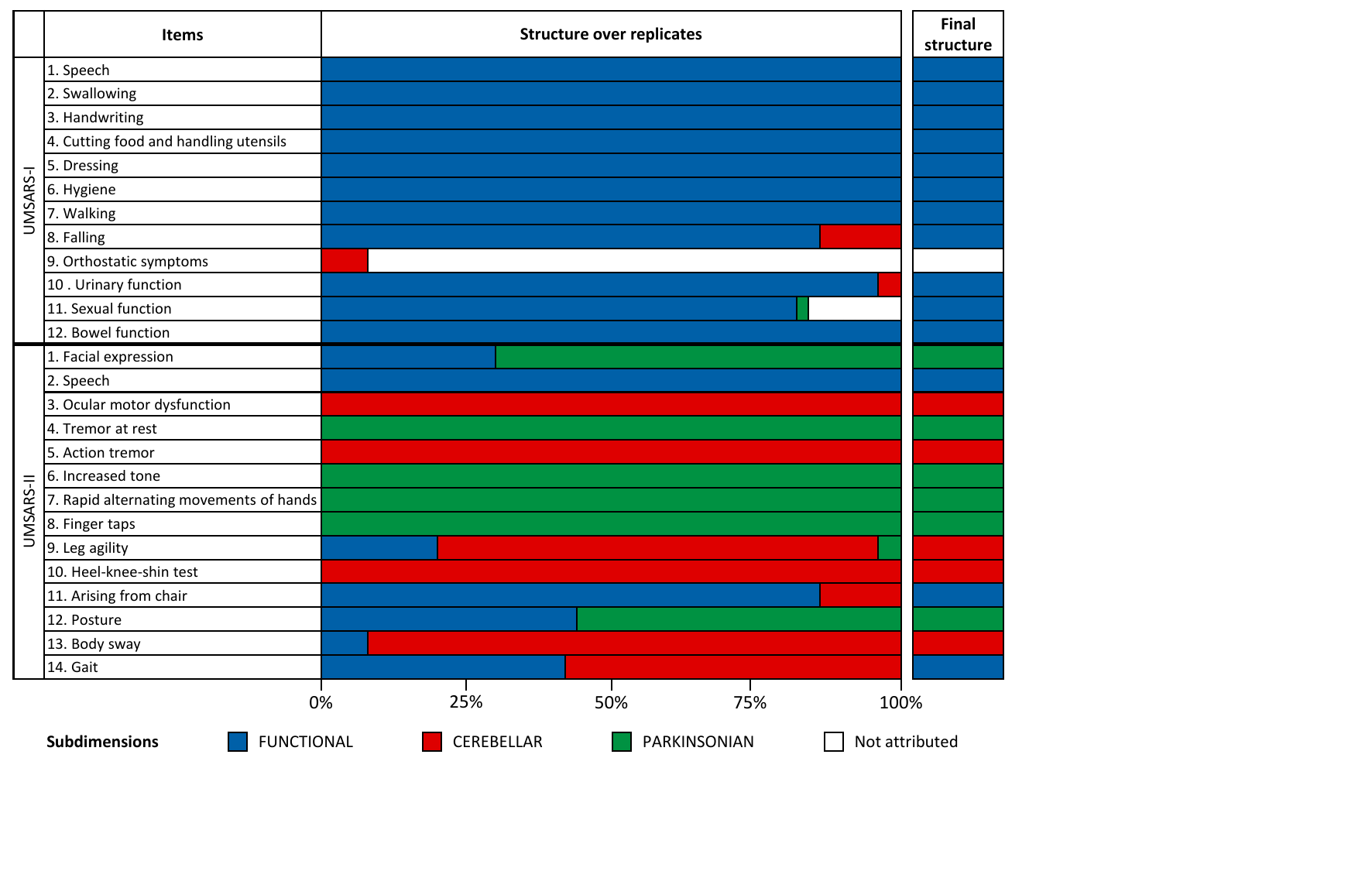}
    \label{tab:msa_umsars}
\end{figure}

The FMSA cohort is the prospective open cohort of the national reference MSA centre located at Bordeaux and Toulouse university hospitals \cite{foubert-samier_disease_2020}. Initiated in 2007, it enrolls all consenting patients diagnosed with MSA \cite{gilman_second_2008} and followed at these hospitals. 
Patients undergo an annual standardized clinical examination based on the four-part UMSARS \cite{wenning_development_2004} which measures: impairment in daily activities (UMSARS-I), motor function (UMSARS-II), orthostatic hypotension (UMSARS-III), and degree of global disability (UMSARS-IV). \RVW{This study focuses on the 26 items of UMSARS-I and UMSARS-II listed in Figure \ref{tab:msa_umsars} (left column).} Each item rates the impairment severity with five increasing levels (0- no, 1- slight, 2- moderate, 3- marked, and 4- extreme impairment). \RVW{UMSARS-IV defines five MSA disease stages: completely independent (stage I), needs help with some chores (stage II), needs help with half of the chores (stage III), does a few chores alone (stage IV), and totally dependent (stage V).} \\

Factors of progression under study are: sex, age at inclusion, \RVW{subtype (MSA-P for parkinsonian  or MSA-C for cerebellar)}, diagnosis certainty (probable or possible), delay between first symptom onset and inclusion, nature of first symptoms (motor, dysautonomic, or both), and hypotension as a first symptom. The status and exact times of death are continuously updated. \RVW{At administrative censoring (October 24, 2022) patients were classified as deceased, still alive, or having dropped out if their last visit was more than 18 months before}.


\section{The 4S method} \label{s:4S}

\RVW{We consider a questionnaire comprising $K$ ordinal items} with $Y_{i}^k(t_{ij})$ the value of item $k$ ($k$=1,...,$K$) for patient $i$ ($i$=1,...,$N$) at visit $j$ ($j$=1,...,$n_{i}$) and time $t_{ij}$ ($t_{ij} \in \mathbb{R^+}$). \RVW{Item $k$ is rated with $M^k+1$ ordered levels from $0$ to $M^{k}$.} Let $Y^S_{i}(t_{ij})$ denote the disease stage with values from $1$ to $S$. For the sake of readability, we consider a common vector of $n_i$ times although the 4S method handles item/stage-specific measurement times. We define $T_i$ the minimum between the time of first event and the censoring time ($T_i \ge t_{in_i}$), and the indicator $d_i=p$ if the patient experienced first the event of cause $p$ ($p$=1,...,$P$) and $d_i=0$ if censored.\\

The four steps of the 4S method, schematized in Figure \ref{fig:4Smethod}, are detailed in the next subsections. Since each step may rely on slightly different information, the samples may vary across steps. The specific requirements and selections are detailed in the corresponding subsections.


\subsection{{\bf Step 1 - Structuring:} \RVW{Identification of the dimensions of the questionnaire}} \label{ss:4S1}

The \RVW{classical IRT methodology usually relies on three assumptions:}  {\bf unidimensionality} (items measuring the same homogeneous latent trait should be analyzed together), {\bf conditional independence} (the latent trait should capture the whole correlation across items with no remaining residual correlation between items) and {\bf increasing monotonicity} (higher item level should consistently reflect a higher underlying trait level) \cite{reeve_psychometric_2007}. \\

\RVW{The first preliminary step consists in structuring the questionnaire into unidimensional traits (interpreted as domains) that comply with these assumptions.} We adapted the recommended strategy reported by the PROMIS (Patient-Reported Outcomes Measurement Information System) initiative for cross-sectional data \cite{reeve_psychometric_2007} to the context of longitudinal data. \RVW{To avoid a distortion of the structure due to the intra-subject correlation, and the varying number of data across subjects, we randomly selected a single visit per subject to create a pseudo independent sample of size $N$ on which the PROMIS method was applied and we replicated this procedure multiple times to account for sampling fluctuations.} \\

The procedure is fully detailed in Appendix A. Briefly, the \textbf{unidimensional} traits are identified by exploratory and confirmatory factor analyses (EFA and CFA) with item assignment based on its highest loading. The final structure is determined after the aggregation over replicates. Items repeatedly not contributing to any dimension are excluded. Items straddling two dimensions are reviewed and assigned based on loadings and clinical relevance. Then the items repeatedly detected with high residual correlation (i.e. lack of \textbf{conditional independence}) over replicates within a dimension are flagged and one is removed based on clinical relevance. The 
\textbf{increasing monotonicity} is visually inspected. \\

After this first step, $D$ dimensions corresponding to homogeneous latent traits are identified. Dimension $d$ ($d$=1,...,$D$) is measured by a set of items $\mathcal{K}^d$, with $\mathcal{K}^d \subset \{1,...,K\}$, $\mathcal{K}^d \cap \mathcal{K}^{d^\prime} = \emptyset$ if $d \neq d^{\prime}$ since items are assigned to a single dimension, and $\bigcup\limits_{d=1}^D \mathcal{K}^d$ represents all the retained items. With the main assumptions verified, each dimension can then be separately analyzed using an IRT-based approach.



\subsection{{\bf Step 2 - Sequencing:} description of items’ impairment sequence and predictors}  \label{ss:4S2}

\begin{figure}[h!]
    \caption{\textbf{Schematic diagrams of the structures of joint models} used in Step 2 (\textbf{A.}) and in Step 3 (\textbf{B.}).}
    \centering
    \includegraphics[width = 12cm, trim= 0cm 2cm 8cm 0cm, clip, page = 1]{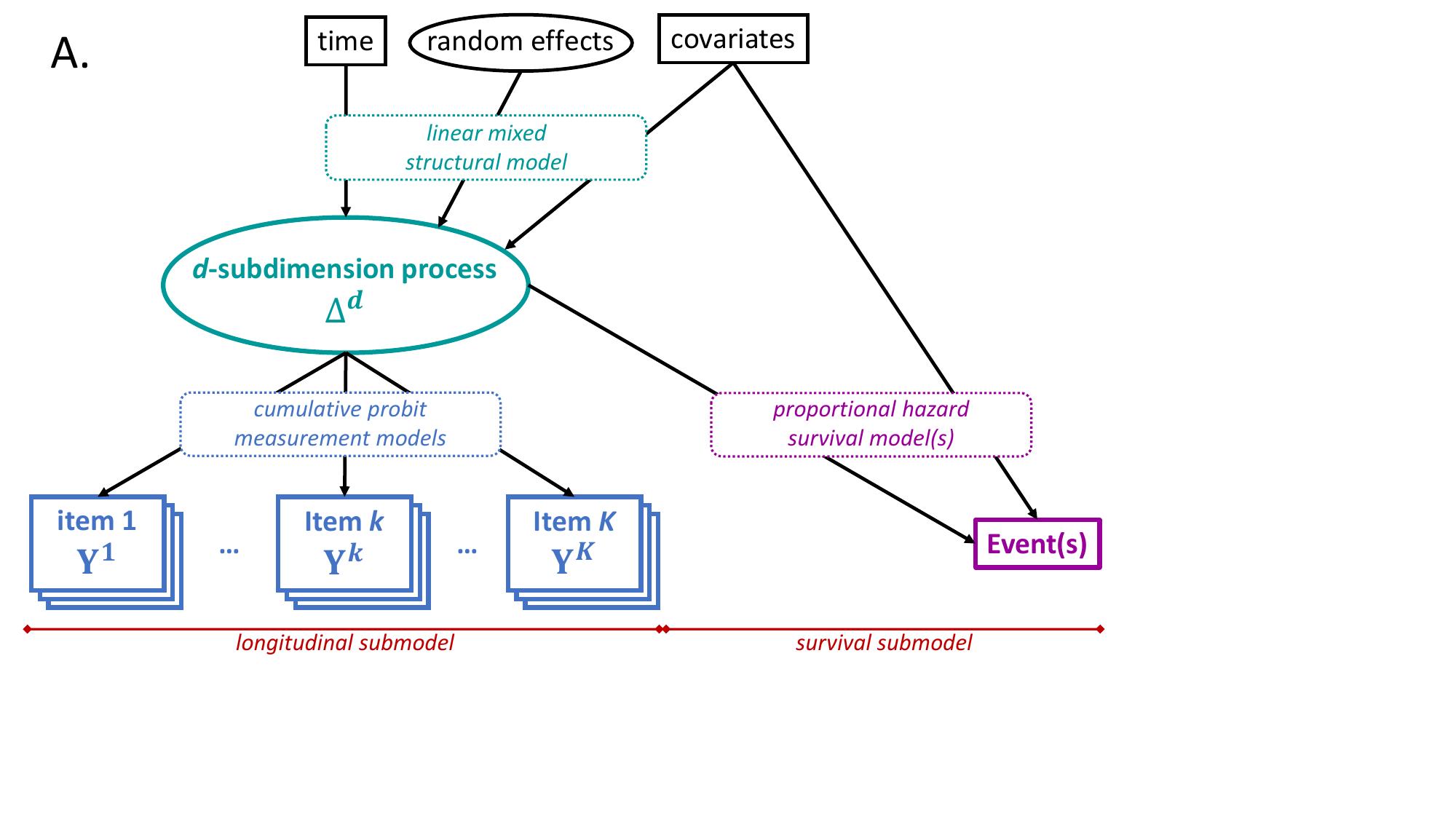}
    \includegraphics[width = 12cm, trim= 0cm 2cm 8cm 0cm, clip, page = 2]{Graphs/ModelSchemes.pdf}
    \label{fig:schema_model}
\end{figure}

For each dimension $d$ ($d=1,...,D$), the value of the underlying latent trait is $\Delta^d_i(t)$ for patient $i$ ($i=1,...,N$) at time $t$ ($t \in \mathbb{R}^+$). Step 2 leverages a joint model combining a continuous-time IRT submodel to describe the items and their underlying trait, and a cause-specific submodel to assess the associated event risk (Figure \ref{fig:schema_model}A) \cite{saulnier_joint_2022}:
\begin{subequations}
    \begin{empheq}[left=\empheqlbrace]{align}
        \Delta^d_i(t) &= \bfs{X^d_i(t)^\top \beta^d} + \bfs{Z^d_i(t)^\top b^d_i} \label{eq:dim}\\
        Y^{k}_{i}(t_{ij}) &= m \Leftrightarrow \delta_{k,m} < \Delta^d_i(t_{ij}) + \epsilon^{k}_{ij} \leq \delta_{k,m+1} ~\text{for}~ k \in \mathcal{K}^d, m=0,...,M^k, j=1,...,n_{i}  \label{eq:item}\\
        \lambda^{dp}_{i}(t) &= \lambda^{dp}_0(t;\bfs{\xi^{dp}}) \exp \left( \bfs{W^{dp\top}_i\gamma^{dp}} + \bfs{g^{dp}(b^d_i, t)^\top\alpha^{dp}} \right) ~\text{for}~ p=1,...,P \label{eq:surv}
    \end{empheq}
\end{subequations}

Equation \eqref{eq:dim} defines a structural linear mixed model for dimension-$d$ trajectory over time according to $\bfs{X^d_i(t)}$ and $\bfs{Z^d_i(t)}$, two vectors of functions of time and \RVW{exogenous covariates}, associated to fixed effects $\bfs{\beta^d}$ and individual normally distributed random effects $\bfs{b^d_i}$ ($\bfs{b^d_i} \sim \mathcal{N}(0, \bfs{B^d})$), respectively. \RVW{The random effects are assumed independent of the exogenous variables}.  $\bfs{X^d_i(t)^\top \beta^d}$ defines the mean trajectory of the dimension at the population level while $\bfs{Z^d_i(t)^\top b^d_i}$ captures the individual deviation to the mean trajectory. 

Equation \eqref{eq:item} defines the measurement model linking each item $k$ ($k \in \mathcal{K}^d$) to latent trait $d$. \RVW{Independent measurement errors} are Gaussian $\epsilon^{k}_{ij} \sim \mathcal{N}(0, \sigma^2_{k})$ with $a_k=1 \slash \sigma_k$ the discrimination parameter representing item-$k$ ability to assess trait-$d$. The measurement models are cumulative probit models with difficulty levels $-\infty= \delta_{k,0} \leq \delta_{k,1} \leq ... \leq \delta_{k,m} \leq ... \leq \delta_{k,M_{k}} \leq \delta_{k,M_{k}+1} = +\infty$. The conditional probability that item $k$ equals level $m$ is:
\vspace{-.25cm}
\begin{equation}
\begin{array}{ll}
\RVW{    P^{k}_m(\Delta) = 
         \Phi\left( a^k (\delta_{k,m+1}-\Delta)\right) - \Phi\left( a^k (\delta_{k,m}-\Delta)\right)}
\end{array}\label{proba_item}
\end{equation}
with $\Phi(.)$ the Gaussian cumulative distribution function. 

Equation \eqref{eq:surv} defines the cause-specific proportional hazard model linking latent trait $d$ with cause-$p$ instantaneous event risk considering a parametric baseline hazard $\lambda^{dp}_0(t;\bfs{\xi^{dp}})$ either piecewise- constant, Weibull or approximated by cubic M-splines. Predictors are entered log-linearly. This includes a covariate vector $\bfs{W^{dp}_i}$ associated to parameters $\bfs{\gamma^{dp}}$ and a function $\bfs{g^{dp}(b^d_i, t)}$ capturing the dimension-$d$ dynamics, either through its random effects $\bfs{g^{dp}(b^d_i, t)} = \bfs{b^d_i}$ or its current value $g^{dp}(b^d_i, t) = \Delta^d_i(t)$, associated to the parameters $\bfs{\alpha^{dp}}$.\\

The maximum likelihood estimation of this Joint Latent Process Model (JLPM) is implemented in the R-package JLPM  \cite{saulnier_joint_2022} with the simultaneous estimation of all the parameters\\ $\bfs{\theta^d}=(\bfs{\beta^d},\bfs{\text{vec}(B^d)},\{ (\delta_{k,m})_{m=1,...,M^k}, \sigma_k \}_{k \in \mathcal{K}^d}, \{ \bfs{\xi^{dp}}, \bfs{\gamma^{dp}}, \bfs{\alpha^{dp}} \}_{p=1,...,P})^\top$ involved in equations \eqref{eq:dim}, \eqref{eq:item}, \eqref{eq:surv} using a Marquardt-Levenberg optimization algorithm, with stringent criteria ensuring convergence \cite{philipps_robust_2021}, and Quasi Monte-Carlo approximation of the integral over the random effects in the log-likelihood computation. \RVW{JLPM assumes missing at random intermittent item data, and dropout possibly missing not at random with mechanisms specific to the cause.}\\

Once each model is estimated (with $\bfs{\hat{\theta}^d}$ the parameter estimates for dimension $d$, and $\bfs{\widehat{V(\hat{\theta}^d)}}$ their variance estimates), the sequence of item impairments along each dimension continuum can be obtained from the estimated difficulty parameters $(\hat{\delta}_{k,m})_{k \in \mathcal{K}^d,m=1,...,M_{k}}$. \RVW{Each threshold $\delta_{k,m+1}$ corresponds to the location in the dimension-$d$ continuum where item-$k$ probability to be lower than level $m$ and item-$k$ probability to be strictly higher than level $m$ are both $0.5$ ($ \mathbb{P}(Y^k(t) \leq m \mid \Delta(t) = \delta_{k,m+1}) = \mathbb{P}(Y^k(t) > m \mid \Delta(t) = \delta_{k,m+1}) = 0.5$). It can thus be interpreted as the expected transition from level $m$ to level $m+1$, and be used to describe the sequence of dimension-$d$ item impairments.} \\

The predicted trajectory of each item $k$ can also be computed from the estimated model:
\begin{equation}
    \widehat{Y^k}(t; \bfs{\theta^d}) = \mathbb{E} \bigr[ Y^k(t); \bfs{\theta^d} ] = \int_{\mathbb{R}^L} \mathbb{E} [ Y^k(t) \mid \Delta^d(t) = \bfs{X^d(t)}^\top \bfs{\beta^d} + \bfs{Z^d(t)}^\top \bfs{b} \bigr] f(\bfs{b}) d\bfs{b}
\end{equation}

where the expectation of $Y^k$ conditional on $\Delta^d(t)$ is:
\begin{equation}
\begin{array}{ll}
    \mathbb{E} [ Y^k(t) \mid \Delta^d(t) = \Delta] &= \displaystyle \sum_{m=0}^{M^k} m \times \mathbb{P}(Y^k(t)=m \mid \Delta^d(t) = \Delta) \\ 
        &= \displaystyle M^k - \sum_{m=0}^{M^k-1} \Phi\left( a^k (\delta_{k,m+1}-\Delta) \right)
\end{array}
\end{equation}

\noindent Note that the integral over the $L$-vector of random effects $\bfs{b}$ can be approximated by Quasi Monte-Carlo algorithm, and that the mean predicted trajectory of each item can be estimated at the point estimate $\bfs{\hat{\theta}^d}$, and its posterior distribution approximated by parametric bootstrap drawing a large number $B$ of $\bfs{\theta^d_b} \sim \mathcal{N}(\bfs{\hat{\theta}^d}, \bfs{\hat{V}(\hat{\theta}^d)})$ ($b=1,...,B$).



\subsection{{\bf Step 3 - Staging:} Projection of disease stages onto the item impairment hierarchy} \label{ss:4S3}

Given that dimension trajectories are independently described in Step 2, the impairment sequence cannot be compared across dimensions. Step 3 aims to align each dimension's continuum with the course of the disease, summarized by clinical stages, thus providing a temporal clinical anchoring. This is achieved through two substeps: we anchor a proxy of each dimension (the sum-score of the $\mathcal{K}^d$-items, $\bar{Y}^d_i(t_{ij}) = \sum_{k \in \mathcal{K}^d} Y_i^k(t_{ij})$) to the clinical stages $Y^S_i(t_{ij})$, and we project this anchoring onto the dimension continuum $\Delta_i^d$.\\

First, the repeated measures of $Y^S_i$ and $\bar{Y}^d_i$ are modeled using a JLPM (Figure \ref{fig:schema_model}B). Structurally similar to the one of Step 2, it assumes a common latent process $\Omega^d(t)$ with:
\begin{subequations}
    \begin{empheq}[left=\empheqlbrace]{align}
        \Omega^d_i(t) &= \bfs{X_i(t)}^\top \bfs{\mu^d} + \bfs{Z_i(t)}^\top \bfs{v^d_i} \label{eq:fact} \\
        Y^S_i(t_{ij}) &= s ~\Leftrightarrow~ \omega^{d}_s < \Omega^d_i(t_{ij}) + \varepsilon^{Y^Sd}_{ij} \leq \omega^{d}_{s+1} ~~ \text{ for } s \in \{1,...,S\} \label{eq:stage} \\
        \bar{Y}^d_i(t_{ij}) &= {H_d}^{-1}\left( \Omega^d_i(t_{ij}) + \varepsilon^{\bar{Y}d}_{ij}\right) \label{eq:score} \\
        \zeta^{dp}_{i}(t) &= \zeta^{dp}_0(t;\bfs{\psi^{d}}) \exp \left( \bfs{W^{dp\top}_i}\bfs{\kappa^{dp}} + \bfs{g^{dp}(v^d_i, t)}^\top\bfs{\tau^{dp}} \right) \label{eq:surv_stage}
    \end{empheq}
\end{subequations}

Equation \eqref{eq:fact} describes the trajectory of the common factor of $Y^S_i$ and $\bar{Y}^d_i$ over time $t$ ($t \in \mathbb{R}^+$) according to covariates (mainly time functions) using the same structure of linear mixed model as in equation \eqref{eq:dim}, with random effects $\bfs{v_i^d} \sim \mathcal{N}(0, \bfs{V^d})$. For sake of simplicity, we did not change the notations for exogenous $\bfs{X}$ and $\bfs{Z}$ although they can differ. 

Equations \eqref{eq:stage} and \eqref{eq:score} define the measurement models for the clinical stages and the sum-scores, respectively. For the ordinal clinical stage $Y^S_i$, a cumulative probit model is assumed as in Step 2 with Gaussian independent measurement error $\varepsilon^{Y^Sd}_{ij}$ and  thresholds $(\omega^{d}_{s})_{s=2,...,S}$. For the sum-score $\bar{Y}^d_i$, we consider a curvilinear measurement model with normally distributed measurement error $\varepsilon^{\bar{Y}d}_{ij} \sim \mathcal{N}(0, \sigma^2_{\bar{Y}d})$. Since the sum-score is a sum of ordinal items, it is likely subject to unequal interval scaling which can be handled using a nonlinear parameterized continuous link function $H_d(.)$ (quadratic I-splines here) \cite{proust-lima_analysis_2013}.

Equation \eqref{eq:surv_stage} simultaneously models the  events according to the common factor dynamics to account for possible informative truncation of the longitudinal data as in Step 2. \\

Once the model is estimated, the threshold parameters $(\hat{\omega}^{d}_s)_{s=2,...,S}$ give the sequence of stage transitions along continuum $\Omega^d$. For transition $(s-1) \rightarrow s$ ($s=2,...,S$), the correspondence in terms of sum-score is given by the conditional expectation $\mathbb{E} \left[ \bar{Y}^d \mid \Omega^{d} = \hat{\omega}^{d}_s \right] = \mathbb{E} \left [ H_d^{-1} (\hat{\omega}^{d}_s + \varepsilon^{\bar{Y}d}) \right]$ that can be computed by Monte Carlo approximation with $\varepsilon^{\bar{Y}d}$ the Gaussian error \cite{proust-lima_analysis_2013}.
Note that all the expectations of variables conditional on dimensions are time-independent so we omit time here for sake of readability.\\

Second, since sum-score $\bar{Y}^d$ can also be predicted by items $Y^k$ ($k \in \mathcal{K}^d$), the threshold parameters $(\hat{\omega}^{d}_s)_{s=2,...,S}$ can be projected on continuum $\Delta^d$ by searching $(\hat{\delta}^{d}_s)_{s=2,...,S}$ such that:

\begin{equation}
    \mathbb{E} \left[ \sum_{k \in \mathcal{K}^d} Y^k \mid \Delta^d = \hat{\delta}^d_s \right] = \mathbb{E} \left[\bar{Y}^d \mid \Omega^{d} = \hat{\omega}^d_s \right]
    \label{eq:stage_location}
\end{equation}

\begin{equation}
\begin{split}
\text{with } \mathbb{E} \left[ \sum_{k \in \mathcal{K}^d} Y^k \mid \Delta^d = \delta^d_s \right] & =  \sum_{k \in \mathcal{K}^d} \mathbb{E} \left[ Y^k \mid \Delta^d = \delta^d_s \right] \\
&= \sum_{k \in \mathcal{K}^d} \biggl\{ M^{k} - \sum_{m=0}^{M^{k}-1} \Phi \left ( a_k(\delta_{k,m+1} - \delta^d_s) \right )\biggl\}
\end{split}
\end{equation}

For each dimension, the estimated stage transition parameters $(\hat{\delta}^d_s)_{s=2,...,S}$ are on the same continuum $\Delta^d$ as the impairment threshold $(\hat{\delta}_{k,m})_{m=1,...,M_{k}-1}$ of item-$k$ ($k \in \mathcal{K}^d$). They can thus serve as a clinical anchor for item impairment description, as illustrated in Figure \ref{fig:4Smethod}. By convention, $\delta^d_{1}=-\infty$ and $\delta^d_{S+1}=+\infty$.



\subsection{{\bf Step 4 - Selecting:} Selection of the most informative items by disease stage}  \label{ss:4S4} 
With Step 4, we aim to identify what are the dominating items of each dimension at each clinical stage to ultimately select the most informative ones for clinical monitoring. \RVW{We quantify the contribution of item $k$ ($k \in\mathcal{K}^d$) along dimension-$d$ continuum ($d$=1,...,$D$) using the Fisher information function \cite{magis_note_2015}. Under correct model specification, the Fisher information is given by the negative expected second derivative of the item probability, and can be written  \cite{baker_item_2004}}: 
\begin{equation}    
    \label{eq:fisher_fct}
    I^{k}(\Delta) = - \mathbb{E} \left[ \frac{\partial^2}{\partial\Delta^2} \log \mathbb{P}(Y^{k} \mid \Delta) \right] = - \sum_{m=0}^{M^{k}} P^{k}_m(\Delta) \frac{\partial^2}{\partial\Delta^2} \log P^{k}_m(\Delta)
\end{equation}
where $\mathbb{P}(Y^{k} \mid \Delta) = \displaystyle \prod_{m=0}^{M^k} P^{k}_m(\Delta)^{\mathds{1}_{Y^k=m}}$ is the item probability at dimension level $\Delta$ with $ P^{k}_m(\Delta)$ defined in Equation \eqref{proba_item}.\\

\RVW{We estimate the total information carried by an item by integrating the information $I^{k}(\Delta)$ computed at the parameter estimates over $\Delta$. Similarly, we estimate the stage-$s$-specific information carried by item $k$ by integration over the interval $[\hat\delta^d_s;~\hat\delta^d_{s+1}]$ that correspond to the transition thresholds for clinical stage-$s$ estimated in Step 3 (Equation \eqref{eq:stage_location}):}
\begin{equation}
\begin{array}{ll}
    I^k_s &= \displaystyle \int_{\hat\delta^d_s}^{\hat\delta^d_{s+1}} I^{k}(\Delta) d\Delta= \displaystyle \sum_{m=0}^{M^{k}} \biggl\{ \int_{\hat\delta^d_s}^{\hat\delta^d_{s+1}} \frac{P^{k\prime}_m(\Delta)^2}{P^{k}_m(\Delta)} d\Delta - \left( P^{k\prime}_m(\hat\delta^d_{s+1}) - P^{k\prime}_m(\hat\delta^d_s) \right)  \biggl\} 
      \label{eq:fisher_stage} 
\end{array}
\end{equation}  

with $P^{k\prime}_m(\Delta) = \displaystyle \frac{\partial}{\partial \Delta} P^{k}_m(\Delta) = -\hat a^k \phi\left(\hat a^k(\hat\delta_{k,m+1}-\Delta)\right) + \hat a^k \phi\left(\hat a^k(\hat\delta_{k,m}-\Delta)\right)$ and $\phi(.)$ the Gaussian density function.\\

In fine, by denoting $\displaystyle I_s = \sum_{k \in \mathcal{K}^d} I^{k}_s$ the total information carried by the items of dimension $d$ at stage $s$, the percentage of stage-$s$ information carried by each item-$k$, $\displaystyle I_s^{k\%} = \frac{I^{k}_s}{I_s}\times100$, can be used to rank the items within each stage, and to identify the most informative items across disease stages, as illustrated in Figure \ref{fig:4Smethod}.

\subsection{\RVW{{\bf Uncertainty assessment at Steps 2 to 4}}}

\RVW{The variance-covariance matrix of JLPMs estimates was estimated by the inverse of the negative Hessian matrix computed at the point estimates. To propagate this uncertainty in the derived results, we used a parametric bootstrap approach: $1000$ random parameter vectors were drawn from the asymptotic multivariate normal distribution of the JLPMs estimates. We derived $95\%$ confidence intervals for item thresholds, stage projections and item-specific information using the 2.5\% and 97.5\% percentiles.}


\section{Application to MSA progression through UMSARS-I and -II items} \label{s:appli}

\RVW{We illustrate the 4S method to describe the progression of MSA through the 26 items of UMSARS-I and -II repeatedly collected in the FMSA cohort until censoring or death.}

\subsection{Demographics} \label{s:appli:demo}

The full analytical sample, leveraged in Step 1, comprised all the visits with at least one of the 26 items filled out, resulting in 2262 observations for 731 patients. The observations with missing covariates listed in Section \ref{s:cohort} and observations without at least one item filled per dimension identified from Step 1 were removed in Step 2, resulting in 2237 observations for 726 patients. For Step 3, we additionally removed observations with missing global disability (UMSARS-IV) and without at least 75\% of the items collected per identified dimension (threshold chosen to ensure the consistency of the sum-score), resulting in 2164 observations for 719 patients. Step 4 did not involve any further sample selection. See Figure S1 for the sequential selection of the analytical samples at each step. \\

Among the 726 patients in Step 2, the mean length of follow-up was 2.0 years (1.9, 1.5, 2.3 years among deceased, dropped out and administratively censored, respectively) for a maximum follow-up of 12.8 years, and a mean of 3.1 visits per patient (Table S1). The sample was balanced in sex (47.5\% male) and hospital (52.6\% in Bordeaux). On average, patients were 60.6 years old at symptom onset and 65.0 years old upon cohort entry, with an average delay since symptom onset around 4.4 years at entry. Patients were mostly diagnosed with MSA-P (67.9\%) and probable certainty (76.6\%). 
At cohort entry, patients were already affected, scoring on average 21.6 out of 48 on the UMSARS-I subscale and 23.6 out of 56 on the UMSARS-II subscale. Most patients were at early disease stages with 19.7\% at stage I, 43.7\% at stage II, 19.1\% at stage III, 16.1\% at stage IV, and 1.4\% at stage V. During follow-up, 452 (62.3\%) patients died and 105 (14.4\%) dropped out of the cohort.


\subsection{Step 1: Identification of UMSARS-I and -II dimensions}  \label{s:appli:S1}

The structuring into dimensions was achieved by applying the PROMIS methodology on 50 random subsamples of 731 observations each. The 50 replicated EFAs consistently led to a structure of 3 clinically meaningful domains (Figure \ref{tab:msa_umsars}) with a functional domain assessed by 14 items primarily from the UMSARS-I subscale, and two motor domains, one seemingly capturing symptoms predominantly linked to cerebellar dysfunction and the other to parkinsonism, with 5 and 6 items from the UMSARS-II subscale, respectively. Most items consistently contributed to the same dimension across replicates, with exceptions for items II.1 (facial expression), II.12. (posture), and II.14. (gait). These items were assigned to the most contributing dimension, based on higher mean loadings for the first two and after clinical consultation for the last one. Item I.9 (orthostatic symptoms) was removed as contributing constantly poorly to all dimensions. This final structure into 3 dimensions demonstrated a strong fit in CFA, meeting the PROMIS statistical criteria (Table S2). All items passed the inspections of conditional independence and monotonicity (see Appendix B).


\subsection{Step 2: Item trajectories and sequences, and modulating factors}  \label{s:appli:S2}

\subsubsection{JLPM specification} Each domain trajectory was modeled using a basis of natural cubic splines with knots placed at 0, 33\%, 66\% and 95\% of visit times (0, 0.367, 1.993, 5.75 years). This specification, that included a random effect on each spline, was selected by Akaike Information Criterion (AIC) among different alternatives (linear, quadratic, splines with knots at different locations). Trajectories were regressed with simple effects of all covariates. For the functional domain, an interaction effect between sex and subtype was added. The competing risks of dropout and death-prior-to-dropout were simultaneously described according to the current value of the dimension process assuming cubic M-splines (with knots at quantiles of event times (0, 1, 2.271, 3.963, and 12.77 years)) and Weibull baseline risk functions, respectively. The JLPM fixed effect estimates are reported in Table S3 \RVW{and the model goodness-of-fit checking is detailed in Appendix B. As sensitivity analyses, we also estimated the model considering missing at random dropout (i.e., fixing the corresponding association parameter to 0), it led to the same estimates (Table S4).}

\subsubsection{Predicted trajectories} 
\begin{figure}[h!]
    \caption{\textbf{Progression of UMSARS-I and -II items over time} predicted by the JLPMs per dimension (functional, cerebellar, parkinsonian): mean \RVW{(and 95\% confidence interval)} item trajectories over years in the cohort (panel \textbf{A.}) and spider plot of item impairment over years (panel \textbf{B.}). In panel B, the darker the color the later the time in the study, and the more central the higher the impairment. Predictions are computed for the reference profile: a male patient, diagnosed with probable MSA-P, aged 65 years old and without delay since symptom onset at inclusion, with motor impairment as first symptoms and no orthostatic hypotension at disease onset.}
    \centering
    \includegraphics[width = 14cm, trim= 0cm 0cm 0cm 0cm, clip]{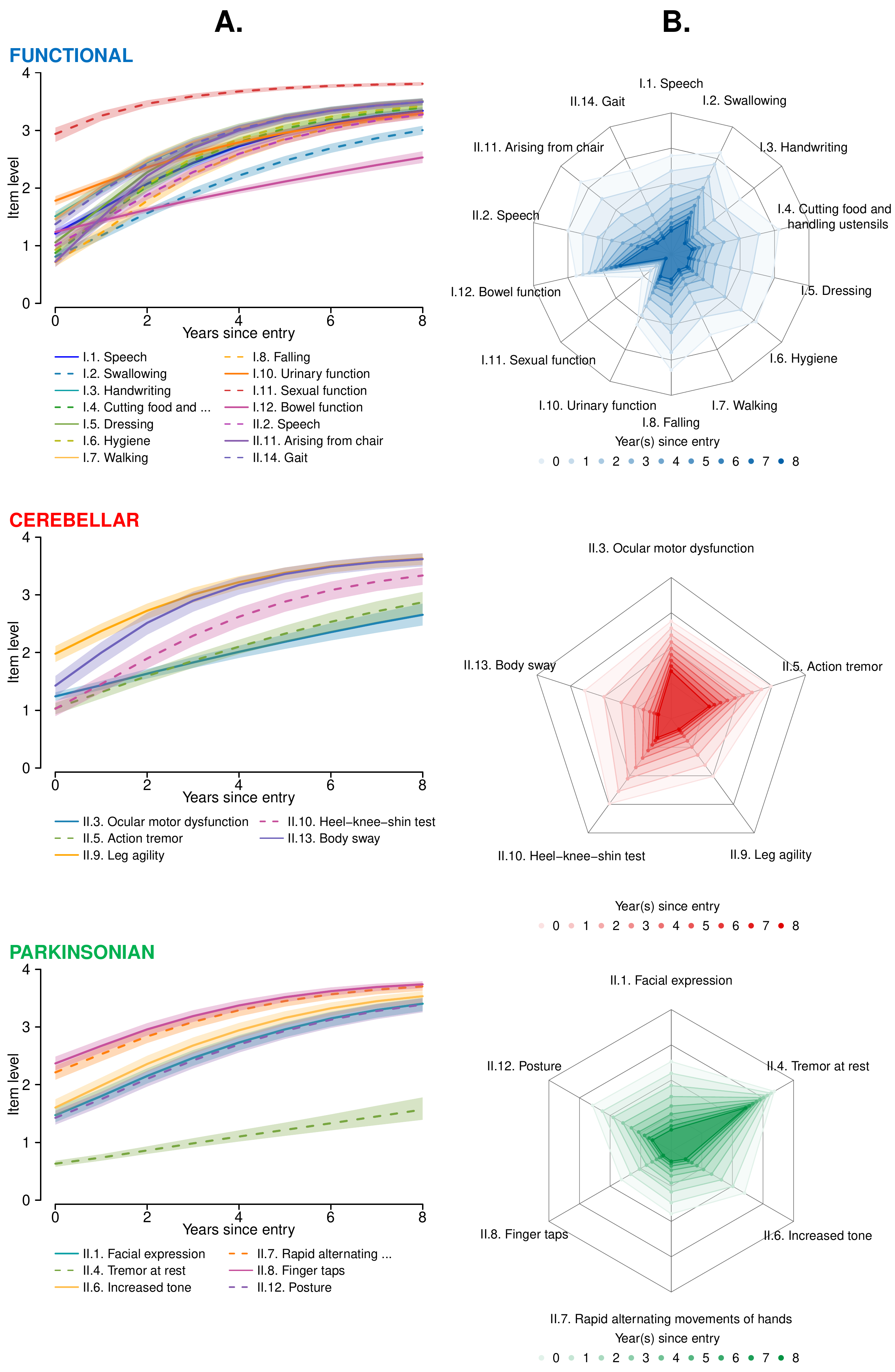}
    \label{fig:msa_itemtraj}
\end{figure}

The predicted items progression are plotted by dimension in Figure \ref{fig:msa_itemtraj} (with trajectories in panel A and spider plots in panel B) \RVW{and Figure S3}. The three domains are already partially affected at entry, with all item mean levels around or above level 1 (first degree of impairment) for the reference profile. All the items of functional domain rapidly deteriorate over the 4 first years of follow-up with exception for item I.11 (sexual function) which is already highly impaired and item I.12 (bowel function) which displays a slower degradation compared to others, surpassing level 3 after 5 or 6 years since entry. In the cerebellar domain, items II.9 (leg agility), II.10 (heel-knee-shin test) and II.13 (body sway) progress rapidly reaching level 3 within 4 to 6 years after entry. In contrast, items II.3 (ocular motor dysfunction) and II.5 (action tremor) exhibit slower progression. For the parkinsonian domain, except for item II.4 (tremor at rest) which starts below level 1 and evolves very slowly (not reaching level 2 after 8 years), all the items begin in mean with substantial initial impairment and progress rapidly. 

\subsubsection{Impairment sequences and modulating factors} 
\begin{figure}[h!]
    \caption{\textbf{Sequence of item impairments according to MSA stages for each dimension, and corresponding intensity of covariate association.} Each graph reports the item degradation according to the underlying dimension continuum, along with the projection of the five MSA disease stage transitions. \RVW{Figure S6 further reports the sequences with the 95\% confidence intervals of all the thresholds}. The intensity of effect of each covariate is indicated by the horizontal black bar with the corresponding p-value of the Wald test.}
    \centering
    \includegraphics[width = 16cm, trim= 0cm 40cm 0cm 0cm, clip]{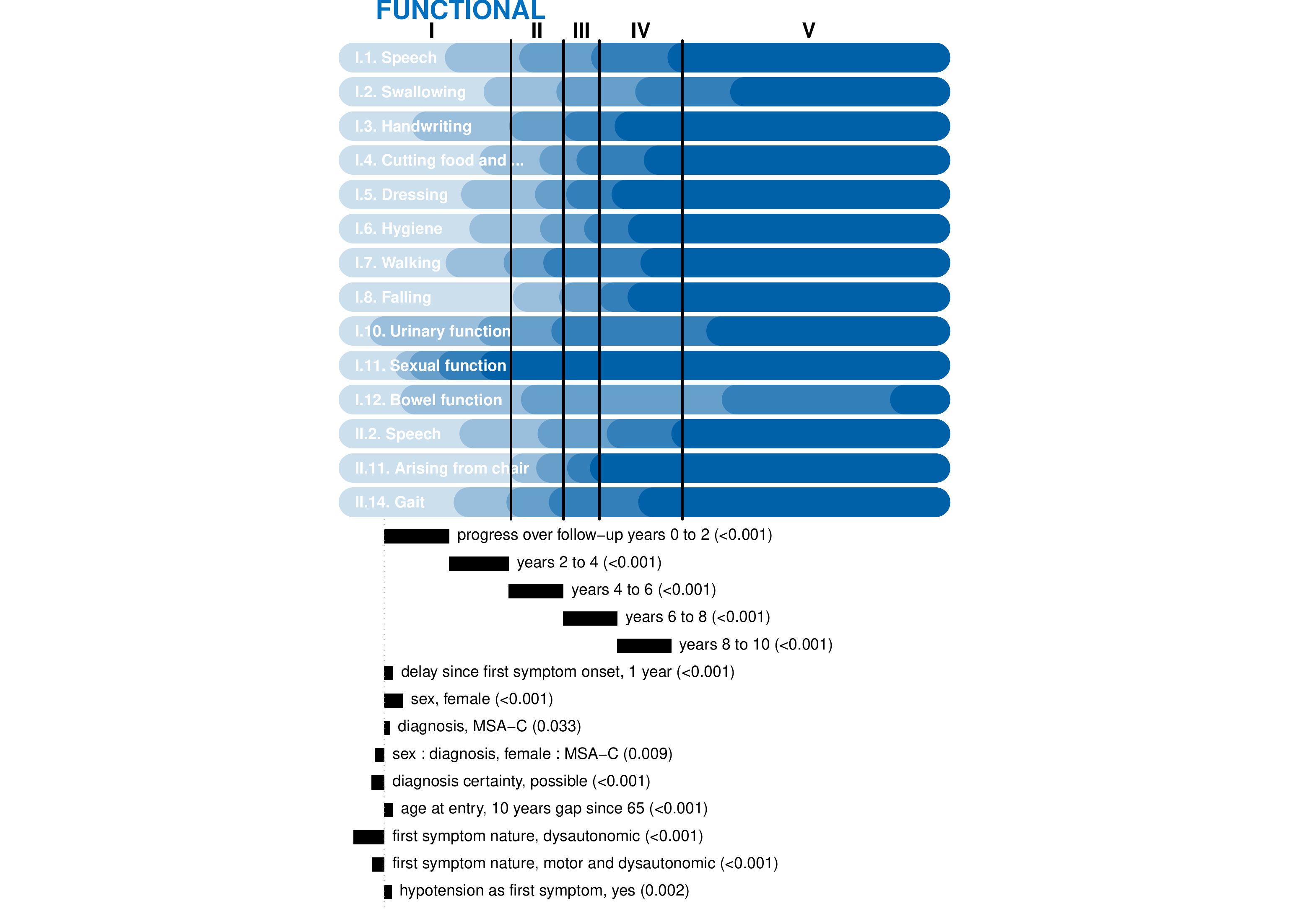}
    \includegraphics[width = 16cm, trim= 0cm 0cm 0cm 47cm, clip]{Graphs/IRT_UMSARS_Hierarchy.pdf}
    \label{fig:msa_hierarchy}
\end{figure}

The sequences of impairments in the three dimensions displayed in Figure \ref{fig:msa_hierarchy} \RVW{(and Figure S6 with 95\% confidence intervals)} confirm these findings. In addition, the dimension levels of impairment differ among patient profiles. As expected, impairment of the cerebellar and parkinsonian domains differ significantly between MSA subtypes, with aggravated impairment of the cerebellar domain for MSA-C patients ($p<0.001$) and of the parkinsonian domain for MSA-P patients ($p<0.001$) (see Figure S3 for sex and subtype differences).
Overall, dimensions' impairment is more severe in patients with longer delay since symptom onset, with probable MSA diagnosis, females and older age. Patients diagnosed solely with motor symptoms also appeared more impaired compared to patients diagnosed with dysautonomic symptoms or both.


\subsection{Step 3: anchoring on MSA disease stages }  \label{s:appli:S3} 

For each dimension, the latent process underlying the domain sum-score and MSA stage was modeled with a natural cubic spline function of time with knots placed at 0\%, 25\%, 50\%, 75\% and 95\% of the visit times (0, 0.805, 1.591, 3.047, 5.75 years). The competing risks of dropout and death-prior-to-dropout were modeled similarly as in Step 2. No further adjustment on covariate was considered. \RVW{The model goodness-of-fit was checked in Appendix B.} \\

The projection of the stages on each dimension continuum (Figure \ref{fig:msa_hierarchy}, Figure S6) revealed that by the end of MSA stage I, most items have transitioned in mean from no impairment (level 0) to slightly impaired (level 1) except items I.8 (falling), II.11 (arising from chair), and II.4 (tremor at rest). With exception for item I.11 (sexual function) that has already reached the maximum, most items then gradually impair during stages II, III, IV with the maximum level reached by the end of stage IV. Only item I.12 (bowel function) and item II.4 (tremor at rest) continue to substantially progress at final stage V.


\subsection{Step 4: the most informative items over MSA course}  \label{s:appli:S4}

\begin{figure}[h!]
    \caption{\textbf{Percentage of information carried by the items of each dimension in the five MSA disease stages}. The total information in each MSA disease stage is reported with a grey bar plot. \RVW{Mean along with 95\% confidence intervals over 1000 bootstrap draws are reported in Table S6.}}
    \centering
    \includegraphics[width = 8.5cm, trim= 0cm 0cm 0cm 0cm, clip]{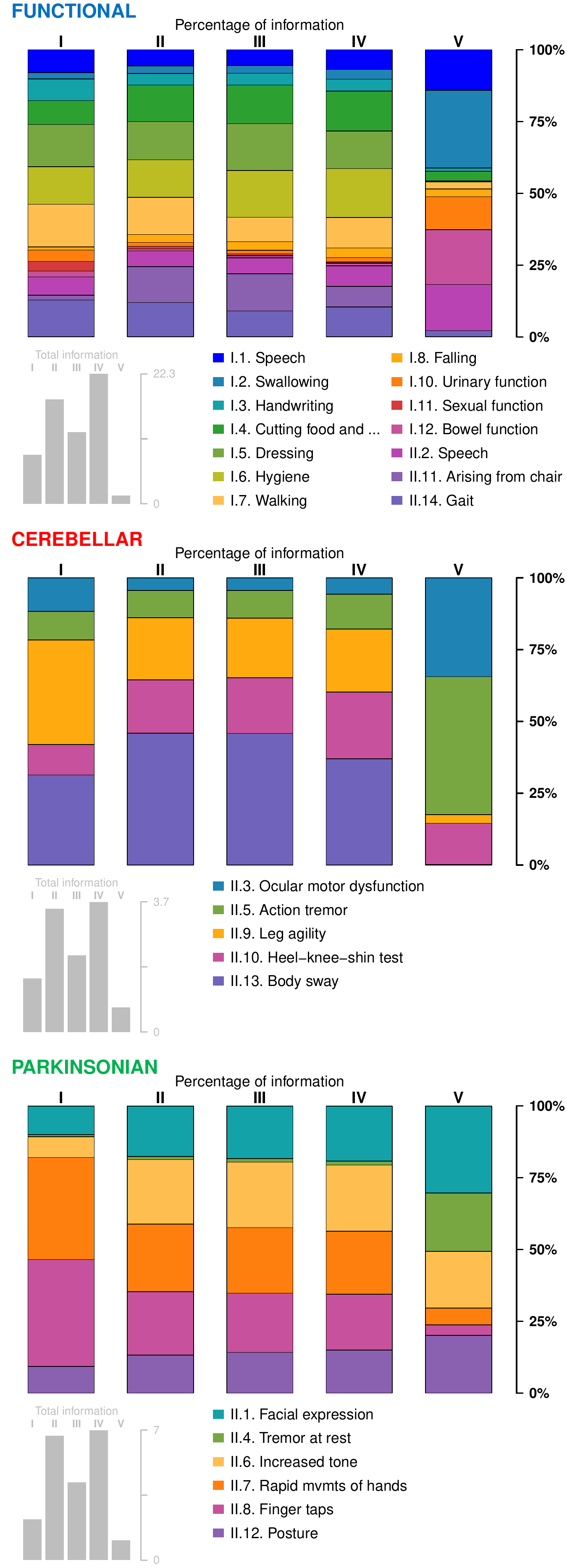}
    \label{fig:msa_fisher}
\end{figure}

The percentages of dimension information carried by the items at each MSA stage are reported in Figure \ref{fig:msa_fisher} \RVW{(and Tables S5 and S6 for ranks and 95\% confidence intervals)}. \RVW{With most items having reached their maximum by the end of Stage IV, Stage V carried little information. We thus only investigated the most informative items at Stages I to IV.} \\

For the functional domain, items I.4 (cutting food and handling ustensils), I.5 (dressing), I.6 (hygiene), I.7 (walking), and II.14 (gait) carried together more than 60\% of information at stage I. Adding item II.11 (arising from chair), the six items then collectively carried 76.5\%, 76.5\%, and 72.3\% at stages II, III, and IV, respectively. On contrary and as expected from the previous results, items I.10 (urinary function), I.11 (sexual function), and I.12 (bowel function) brought negligible information, each contributing less than 4\%. 
For the cerebellar domain, item II.9 (leg agility) carried highest information at stage I with 36.6\% but across Stages I to IV, item II.13 (body sway) concentrated the largest amount of information with between 31.2\% to 45.8\%. 
For the parkinsonian domain, items II.7 (rapid movements of hands) and II.8 (finger taps) are the most informative items across stages with 72.6\% of combined information at stage I, and over 40\% at stages II to IV. In stages II to IV, these two items plus items II.1 (facial expression) and II.6 (increased tone) contributed roughly equally, around 20\% each. As anticipated, item II.4 (tremor at rest) carried very little information ($<$2\%). 


\section{Discussion} \label{s:discuss}

With the 4S method, we established a comprehensive strategy of analysis of repeated \RVW{Likert scales with the structuration of a questionnaire into subdimensions}, their description over time according to covariates, and their progression along disease stages. \\

In the FMSA cohort, the 4S method highlighted \RVW{three domains of impairments from the UMSARS-I and -II items, with varying progression rates and modulating factors over the disease course. This offers new insights into the relative trajectories of symptom deterioration, thereby informing more precise therapeutic and rehabilitation strategies. The magnitude of information also varied across items and disease stages, providing clinicians with key manifestations to monitor and to potentially use as endpoints in clinical trials.} \\

The analysis of \RVW{surveys} has largely spread in psychometrics pointing the importance to interpret data \RVW{at both the domain and item levels depending on the objective}. \RVW{With repeated health-related surveys in clinical studies,} two main additional challenges are that visits don't necessary occur at the exact same times across patients and that the observation process can be interrupted by events inducing informative dropout. In addition, as latent domain values are not directly interpretable, they need to be anchored according to disease course. The proposed strategy meets these challenges combining psychometric approaches from measure theory with mixed and joint modeling techniques from biometrics. 
With COAs becoming increasingly available, it applies to various health areas with significant potential implications. Examples include clinical assessments in psychiatric disorders for instance, and health-related quality-of-life (Hr-QoL) in most chronic diseases (e.g., cancer). \RVW{The analysis of Hr-QoL in MSA patients led for instance to a mind map for guiding clinicians in the management of QoL-related impairments  \cite{saulnier_patient-perceived_2024}}. \\

The 4S method should however be used with caution keeping some limitations in mind. \RVW{The 4S method is dedicated to ordinal and binary items. Analysis of other types of items (e.g., nominal, continuous, or mixed) would require adaptations}. \RVW{The methodology was also designed for describing and understanding the disease progression through questionnaires of items, not for individually predicting the progression.}
\RVW{Step 1 aims to structure the item information at best for its modeling, conforming to the model assumptions. It hinges on multiple decisions 
and is confined to the items listed in the questionnaire studied.} \RVW{In its current version, the 4S strategy assigns each item to a single dimension with varimax rotation. We notably leave for future research the possibility of assigning an item to multiple dimensions \cite{marsh_exploratory_2014, chen_bifactor_2018}}. \RVW{In Steps 2 and 3, each dimension is modeled using a previoulsy validated JLPM technique \cite{saulnier_joint_2022}.} This parametric approach requires an adequate specification in regards to the data. \RVW{Steps 2 and 3 are carried out independently across domains with the disease stages linking them together. Multidimensional IRT allow the simultaneous analysis of multiple dimensions while capturing their associations \cite{reckase_multidimensional_2009,cai_longitudinal_2021}. However, to the best of our knowledge, multidimensional IRT models have not yet been fully extended to longitudinal (with irregular timings) and joint modeling frameworks.}
\RVW{We relied on a cause-specific model for the intercurrent events as we assumed the mechanisms leading to the two events were likely different, and the model assumed proportional hazards for the covariates and the underlying constructs.} \RVW{Finally, we used a parametric bootstrap technique to account for the parameter estimates uncertainty in Steps 2 to 4. This neglected the potential uncertainty in domains structuration in Step 1. However the replicates in Step 1 highlighted a substantial stability in the item assignments to domains (Figure \ref{tab:msa_umsars}), suggesting limited risks of error propagation.} \\

In conclusion, the 4S method offers a comprehensive strategy for the longitudinal analysis of data derived from COAs using innovative statistical models and various graphical tools for interpretation. The 4S method applies in a large variety of heath contexts with potential implications for disease understanding and management, and therapeutic research.


\section*{Replication script}
\href{https://github.com/VivianePhilipps/JLPM/blob/main/vignettes/script_4Smethod.Rmd}{https://github.com/VivianePhilipps/JLPM/blob/main/vignettes/script\_4Smethod.Rmd}


\newpage

\newpage

\newpage

\newpage

\newpage

\newpage

\singlespacing

\bibliographystyle{abbrv} 
\bibliography{main}


\section*{Acknowledgements}

The authors thank the members of the FMSA research group: Prof. Olivier Rascol, Prof. Anne Pavy-Le-Traon, Dr. David Bendetowicz, Federico Sirna, Dr. Catherine Helmer, and Mélanie Le Goff.
Computer time was provided by the computing facilities MCIA (Mésocentre de Calcul Intensif Aquitain) at the University of Bordeaux and the University of Pau and Pays de l’Adour. The authors thank the French National Research Agency (Project DyMES - ANR-18-C36-0004-01), the Nouvelle-Aquitaine region (Project AAPR2021A-2020-11937310) and the French government in the framework of the PIA3 (“Investment for the future”, project reference 17-EURE-0019) for their financial support that made this work possible. 

\section*{Authorship}

TS, CPL and AFS designed the research project, and formulated and developed the statistical method. TS executed the statistical analysis under supervision of CPL and AFS. WGM, MF and AFS shared the French MSA cohort data and provided clinical interpretation and discussion facing the analysis results. TS, CPL and AFS wrote the first draft. All authors reviewed the results, revised and approved the final version of the manuscript.

\end{document}


\begin{center}
    \large \bf Supplementary Materials \\
    
    Structuring, Sequencing, Staging, Selecting: the 4S method for the longitudinal \\ 
    analysis of multidimensional questionnaires in chronic diseases
\end{center}

\begin{center}
    {\bf Tiphaine Saulnier$^{1,*}$, Wassilios G. Meissner$^{2,3,4}$, Margherita Fabbri$^{5,6}$, \\ Alexandra Foubert-Samier$^\mathbf{\dagger1,2,3,7}$, and Cécile Proust-Lima$^\mathbf{\dagger1,7}$}
\end{center}

\vspace{.25cm}
\small
\noindent$^1$Univ. Bordeaux, Bordeaux Population Health research center, Inserm U1219, Bordeaux, France \\
$^2$CHU Bordeaux, Service de Neurologie des Maladies Neurodégénératives, CRMR AMS, NS-Park/FCRIN Network, Bordeaux, France \\
$^3$Univ. Bordeaux, IMN, CNRS, UMR 5293, Bordeaux, France \\
$^4$Dept. Medicine, Univ. Otago, New Zealand Brain Research Institute, Christchurch, New Zealand \\
$^5$CHU Toulouse, MSA French Reference Center, Toulouse, France \\
$^6$Univ. Toulouse, Dept. Clinical Pharmacology and Neurosciences, CIC-1436, NeuroToul COEN Center, NS-Park/FCRIN Network, Inserm U1048/1214, Toulouse, France \\
$^7$Inserm, CIC1401-EC, Bordeaux, France \\
$^\dagger$These authors contributed equally. \\
*\textit{email: tiphaine.saulnier@u-bordeaux.fr}

\normalsize
\vspace{1cm}


\newpage

\subsection*{\RVW{Appendix A: Additional information on the 4S methodology}}
\addcontentsline{toc}{section}{Appendix A: Additional information on the 4S methodology}

\subsubsection*{\RVW{Details on Step 1 - Structuring}}

This appendix subsection details the criteria recommended in the PROMIS strategy to assess unidimensionality, conditional independence, and increasing monotonicity [18]. The three calibration assumptions can be successively evaluated with stepbacks or systematic reevaluations to measure the impact of certain decisions. \\

\textbf{Unidimensionality:} An exploratory factor analysis (EFA) is performed on all items to identify underlying subdimensions measured by the questionnaire. The optimal number of subdimensions can be determined dressing the scree plot of the successive eigenvalues and chosen either as the largest number of factors with successive eigenvalues greater than value 1 (Kaiser criterion), or identifying the "elbow" where the eigenvalues begin to plateau (Cattell criterion). Items are assigned to the subdimension they contribute the most based on factor loadings resulting from the polychoric correlation matrix\RVW{, with {\it varimax} rotation maximizing within-factor variance (ensuring that each item is strongly associated with its assigned dimension and weakly with others),} and representating the strength of association between the items and the latent traits. If the higher loading of an item is lower than $0.3$, its contribution is considered too insufficient to be assigned to any subdimension. \RVW{In ambiguous cases with nearly equal loadings, item assignment can be guided by considering the dimension most frequently selected across replicates, the median loadings over replicates, or clinical expertise - to ensure that the identified dimensions are clinically meaningful and interpretable.} At this stage, one EFA can be performed per subdimension to ensure only one latent factor is highlighted. A confirmatory factor analysis (CFA) is performed on identified subdimensions to evaluate the model fit. According to Reeve et al. [18], the model fit is considered sufficient if the following criteria are satisfied: comparative fit index (CFI) $> 0.95$, Tucker Lewis index (TLI) $> 0.95$, root means square error of approximation (RMSEA) $< 0.06$, and standardized root mean square residual (SRMR) $< 0.08$.
CFI and TLI assess the fit between the estimated model and a null model assuming no constraint on the structure.
RMSEA and SRMR assess the fit  between the estimated model and the observed data. \\

\textbf{Conditional independence:} The correlation matrix between the CFA predicted values and the observations is reported to identify residual correlation. Reeve et al. [18] consider two items are highly correlated if their residual correlation exceeds $0.2$. In case the two correlated items are assigned to the same subdimension, one of them should be removed, as considered as redundant. \\

\textbf{Increasing monotonicity:} The monotonicity is visually inspected for each item by plotting the mean item levels by decile of the rest-score of the assigned subdimension (i.e., sum-score of all the other items from the subdimension except this one), or the item higher-level probabilities by decile of the rest-score. Reeve et al. [18] recommend to visually check that the curves are consistently increasing or at least constant. 
Given the consistency  of  results  across  replicates,  it’s  acceptable  to  verify  this  assumption  on  a  subset  of replicates only. \\

\noindent Step 1 - Structuring analyses can be performed with R. \RVW{A detailed replication script is available as a vignette within R-package JLPM\\ (\href{https://github.com/VivianePhilipps/JLPM/blob/main/vignettes/script_4Smethod.Rmd}{https://github.com/VivianePhilipps/JLPM/blob/main/vignettes/script\_4Smethod.Rmd}).}


\newpage 

\subsection*{\RVW{Appendix B: Additional information on the application to FMSA}}
\addcontentsline{toc}{section}{Appendix B: Additional information on the application to FMSA}

\subsubsection*{1. Selection and description of the sample}

\begin{figure}[h!]
    \caption{\textbf{Flowchart of MSA sample selection.} 
    Patients without any completed UMSARS-I and UMSARS-II subscales were excluded from the analyses. Depending on the step-specific requirements, the analytical sample varies across the four steps. For \textit{Step 2 - Sequencing}, patients with missing values for at least one covariate of interest (among delay since symptom onset, sex, subtype, diagnosis certainty, age, first symptom nature, and hypotension at disease onset), and observations without at least one item completed per identified subdimension were excluded. For \textit{Step 3 - Staging}, observations without MSA stages (UMSARS-IV) or without at least 75\% of the items completed per identified subdimension were excluded. \textit{Step 4 - Selecting} only relies on model estimations from Steps 2 and 3, not on data directly.}
    \vspace{.5cm}
    \centering
    \includegraphics[width = 9cm, trim= 0cm 0cm 17.3cm 0cm, clip, page = 1]{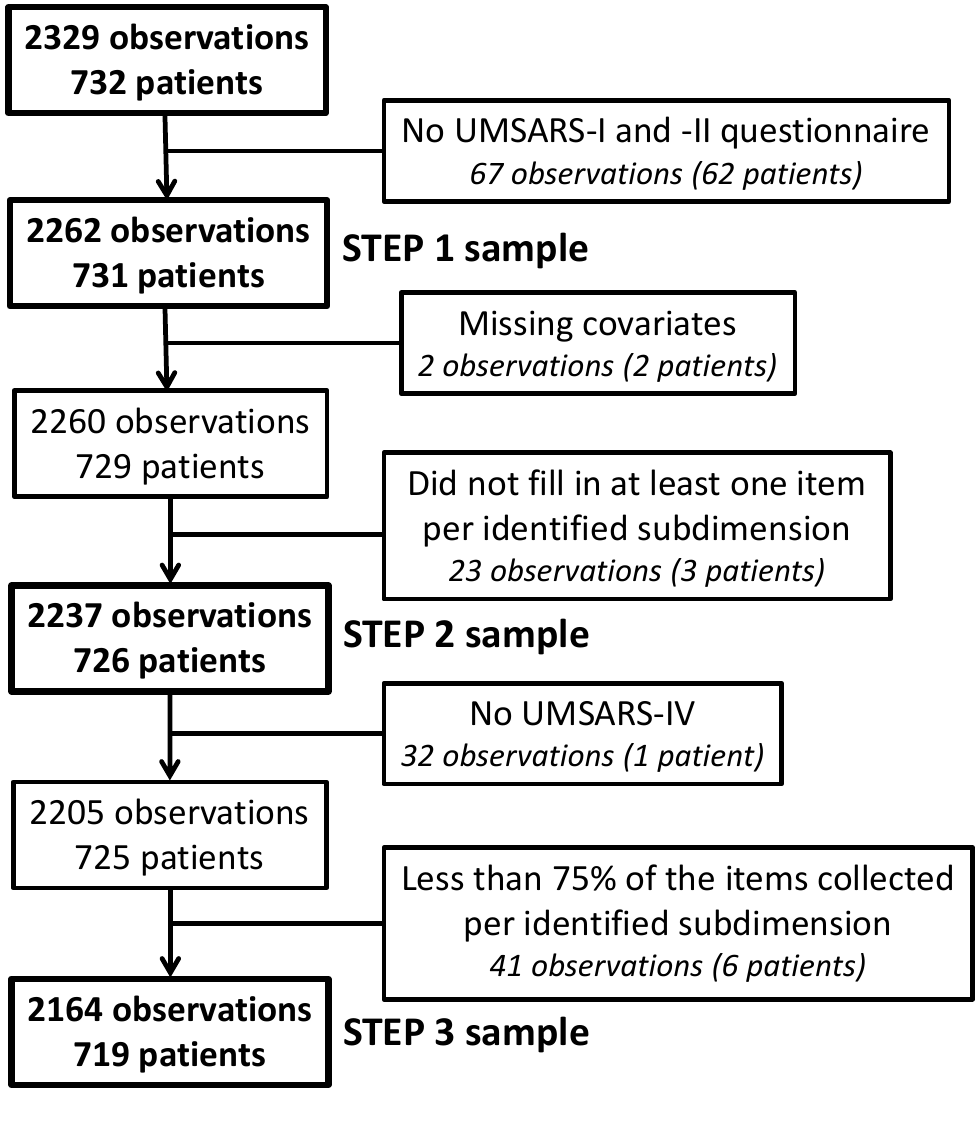}
    \label{fig:msa_flowchart}
\end{figure}

\RVW{Figure \ref{fig:msa_flowchart} reports the flowchart that led to the final analytic samples from FMSA cohort. \\ 
Table \ref{tab:descript} describes the final analytical sample (from Step 2) at inclusion and during follow-up.}

\newpage

\begin{table}[h!]
    \caption{\textbf{Description of the MSA sample (from Step 2 - Sequencing) at inclusion and during follow-up (N=726).}}
    \label{tab:descript}
    \vspace{.5cm}
    \centering
        \begin{tabular}{lcc}
            \hline
            Characteristics & N (\%) & mean $\pm$ sd \\ \hline
            \multicolumn{3}{l}{\textit{At inclusion}} \\
            \hspace{0.25cm} Sex male (vs female) & 345 (47.5) & \\
            \hspace{0.25cm} Centre Bordeaux (vs Toulouse) & 382 (52.6) & \\
            \hspace{0.25cm} Subtype MSA-P (vs MSA-C) & 493 (67.9) & \\
            \hspace{0.25cm} Certainty Probable (vs Possible) & 556 (76.6) & \\
            \multicolumn{3}{l}{\hspace{0.25cm} First symptom nature} \\
                \hspace{0.5cm} Motor & 417 (57.4) & \\
                \hspace{0.5cm} Dysautonomic & 96 (13.2) & \\
                \hspace{0.5cm} Both & 213 (29.3) & \\
            \hspace{0.25cm} Hypotension among first symptoms & 114 (15.7) & \\
            \hspace{0.25cm} Age at symptom onset & & 60.6 $\pm$ 8.3 \\
            \hspace{0.25cm} Age at cohort entry & & 65.0 $\pm$ 8.1 \\
            \hspace{0.25cm} Delay since symptom onset & & 4.4 $\pm$ 2.4 \\
            \multicolumn{3}{l}{\hspace{0.25cm} UMSARS subscales} \\
                \hspace{0.5cm} Sum-score for Part I (out of 48) \textit{(n=698)} & & 21.6 $\pm$ 7.7 \\
                \hspace{0.5cm} Sum-score for Part II (out of 56) \textit{(n=704)} & & 23.6 $\pm$ 8.6 \\
            \multicolumn{3}{l}{\hspace{0.25cm} PROMIS-identified UMSARS-I and -II subdimensions} \\
                \hspace{0.5cm} Sum-score for functional (out of 56) \textit{(n=695)} & & 25.8 $\pm$ 9.8 \\
                \hspace{0.5cm} Sum-score for cerebellar (out of 20) \textit{(n=704)} & & 8.6 $\pm$ 3.3 \\
                \hspace{0.5cm} Sum-score for parkinsonian (out of 24) \textit{(n=706)} & & 9.7 $\pm$ 4.3 \\
            \multicolumn{3}{l}{\hspace{0.25cm} UMSARS Part IV disability degree \textit{(n=702)}} \\
                \hspace{0.5cm} Completely independent (stage I) & 138 (19.7) & \\
                \hspace{0.5cm} Not completely independent (stage II) & 307 (43.7) & \\
                \hspace{0.5cm} More dependent (stage III) & 134 (19.1) & \\
                \hspace{0.5cm} Very dependent (stage IV) & 113 (16.1) & \\
                \hspace{0.5cm} Totally dependent (stage V) & 10 (1.4) & \\
            \hline    
            \multicolumn{3}{l}{\textit{During follow up}} \\ 
            \hspace{0.25cm} Visits & 2237 & \\
            \hspace{0.25cm} Visits per patient & & 3.1 $\pm$ 2.3 \\
            \hspace{0.25cm} Years of follow-up & & 2.0 $\pm$ 2.2 \\
            \hspace{0.25cm} Dropouts & 105 (14.4) & \\
            \hspace{0.25cm} Deaths & 452 (62.3) & \\
            \hline
        \end{tabular}
\end{table}

\newpage

\subsubsection*{\RVW{2. Additional information for Step 1 - Structuring}}

{\bf Unidimensionnality}


\begin{table}[h!]
    \caption{\textbf{CFA fit criteria for MSA application in Step 1 - Structuring} to evaluate the reliability of UMSARS-I and -II structured in functional, cerebellar, and parkinsonian subdimensions (more details about criteria can be found in Appendix A).}
    \label{tab:CFAcrit:MSA}
    \vspace{.25cm}
    \centering
    \small
    \begin{tabular}{lcccc}
        \hline
        Criteria & CFI & TLI & RMSEA & SRMR \\
        Mean value over replicates & 0.989 $\pm$ 0.0017 & 0.988 $\pm$ 0.0019 & 0.099 $\pm$ 0.0075 & 0.065 $\pm$ 0.0039 \\
        Recommended thresholds & \textcolor{colitem4}{$>0.95$} & \textcolor{colitem4}{$>0.95$} & \textcolor{orange}{$<0.06$} & \textcolor{colitem4}{$<0.08$} \\
        \hline
    \end{tabular}
\end{table}

\noindent{\bf Conditional independence} \\

\noindent Residual correlations were observed between item II.6 and items II.3 (-0.231 in mean; $<$-0.200 for 47 replicates) and II.1 (0.234 in mean; $>$0.200 for 15 replicates). Given  these correlations remained moderate and the concerned items were not allocated to the same subdimension, the items were retained in the analyses.\\

\noindent{\bf Increasing monotonicity} \\

\begin{figure}[h!]
    \caption{\textbf{Item higher-level probabilities and mean response according to the allocated subdimension rest-score} for visual checking of the item monotonicity in the randomly selected replicate 49/50.
    The colored curves represent the observed probabilities of response according to the allocated subdimension rest-score, with the gradient color corresponding to increasing level, from the lightest color for the probability of a level equal or higher than 1, to the darkest color for the probability of a level equal to 4. The dotted black curve represents the observed mean level according to the allocated subdimension rest-score.}
    \vspace{.25cm}
    \centering
    \includegraphics[width = 9.5cm, trim= 0cm 0cm 0cm 0cm, clip]{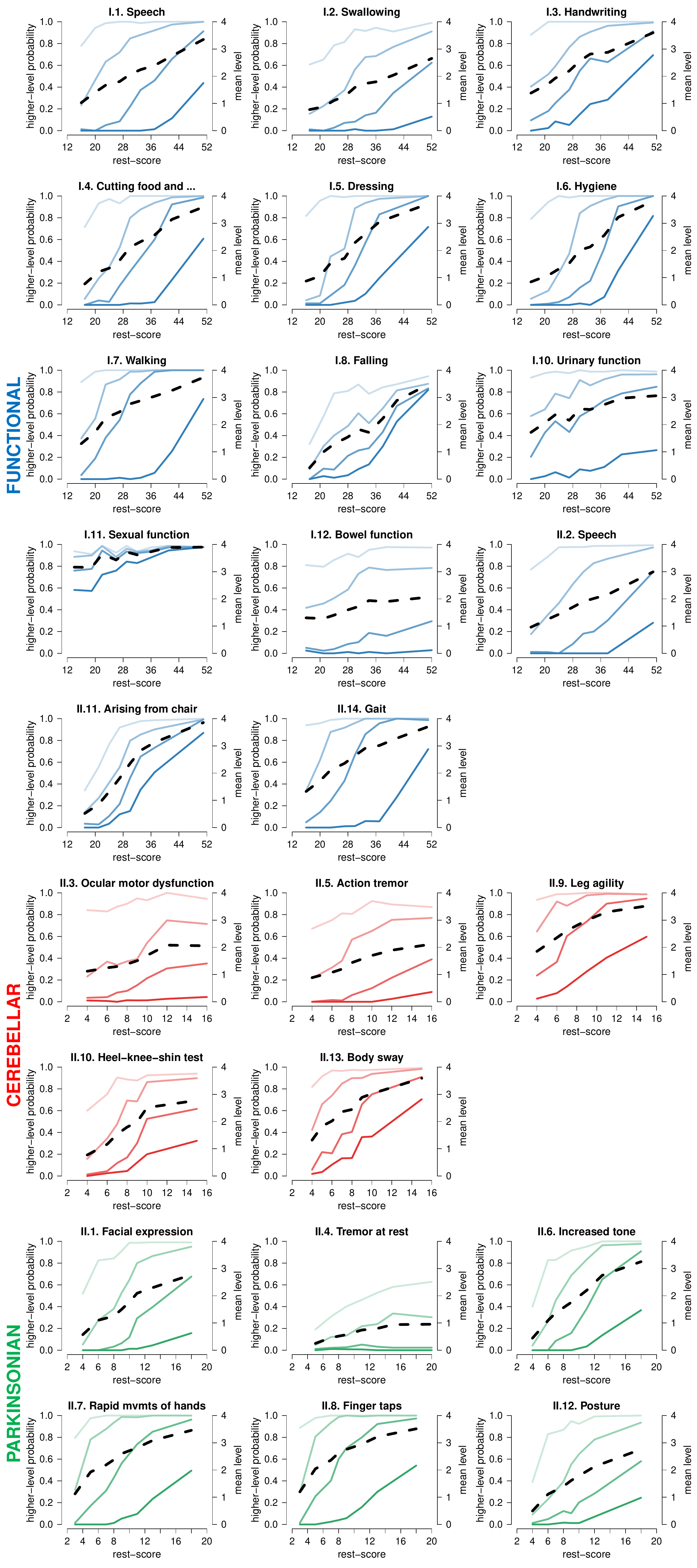}
    \label{fig:msa_promis_checkmono}
\end{figure}

\noindent Visual checking of increasing monotonicity was conducted on some replicates (Figure \ref{fig:msa_promis_checkmono}) and was validated with caution on items I.11 (sexual function) from the functional subdimension as it systematically exhibited high levels, and II.4 (tremor at rest) from the parkinsonian subdimension as it displayed low levels and a modest progression. 

\newpage

\newpage

\subsubsection*{\RVW{3. Additional information for Step 2 - Sequencing}}

\begin{figure}[h!]
    \caption{\textbf{Comparison of the progression 
    of UMSARS-I and -II items according to sex and subtype.} Are reported the mean item levels predicted by a JLPM for each subdimension and at different years from entry. The darker colors indicate later time points and the graph center represent the maximum level of impairment.
    Predictions are displayed in the order: man with MSA-P, man with MSA-C, female with MSA-P, and female with MSA-C. Other factors are fixed to the reference profile: patient aged 65 years old and without delay since symptom onset at inclusion, with motor impairment as first symptoms and no hypotension at disease onset.}
    \vspace{.5cm}
    \centering
    \includegraphics[width = 17cm, trim= 0cm 0cm 0cm 0cm, clip]{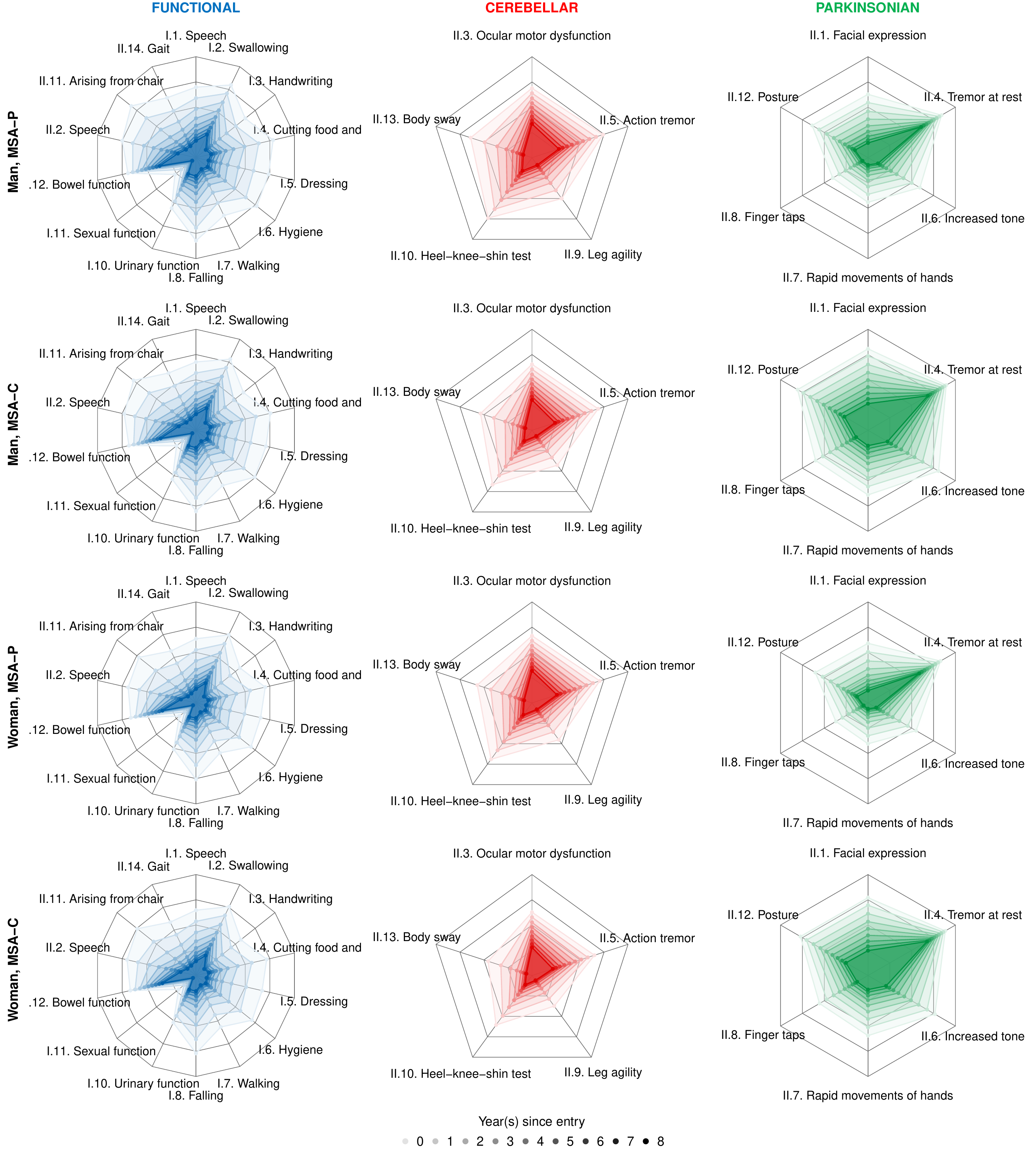}
    \label{fig:msa_spider}
\end{figure}

\begin{figure}[h!]
    \caption{\RVW{\textbf{Diagnostic plots assessing the fit of the longitudinal part in the Step 2 estimated JLPMs.} Graphs compare the observed item means to the predicted item means over a fine grid of times.}}
    \vspace{.5cm}
    \centering
    \includegraphics[width = 14cm, trim= 0cm 0cm 0cm 0cm, clip]{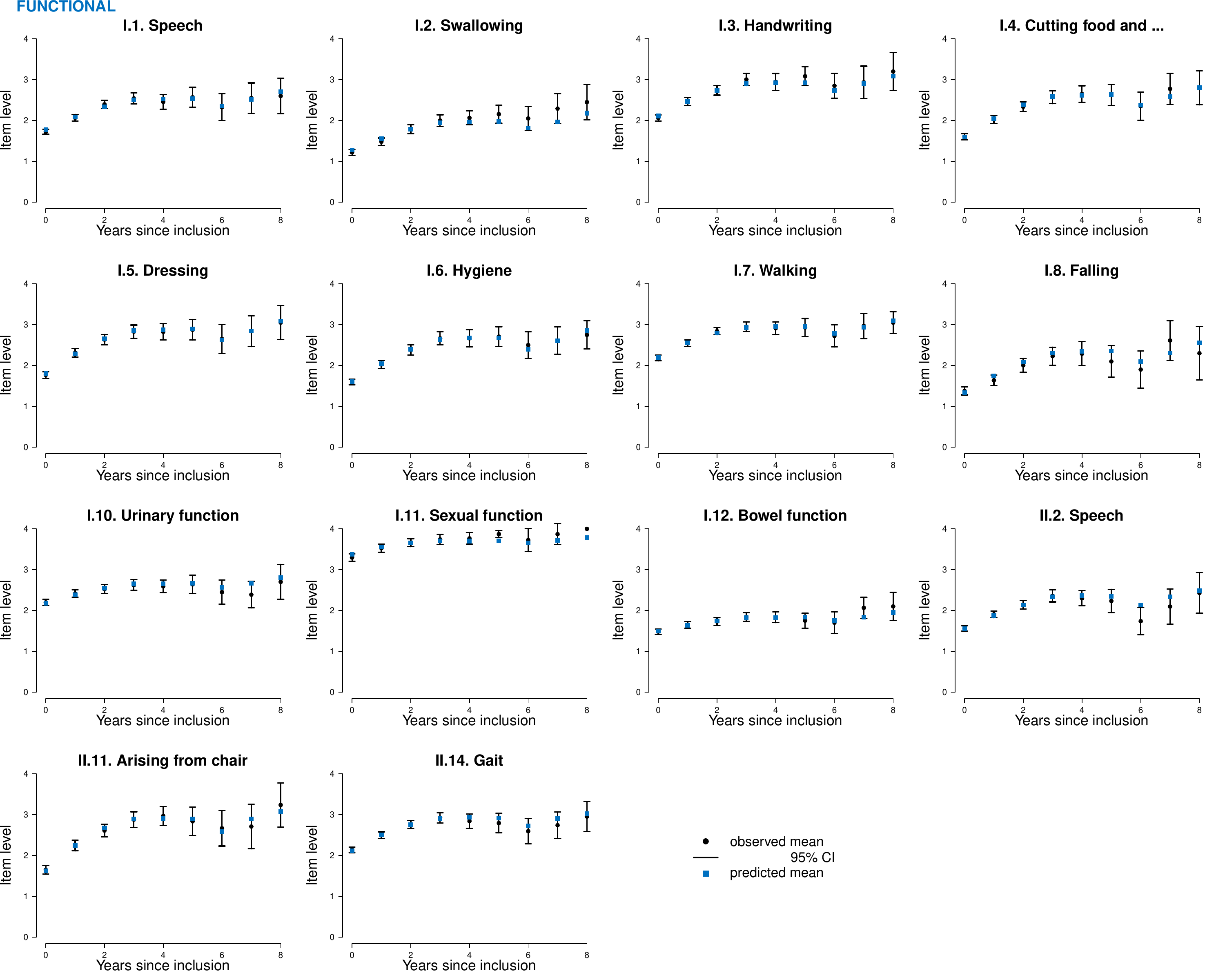}
    \includegraphics[width = 14cm, trim= 0cm 0cm 0cm 0cm, clip]{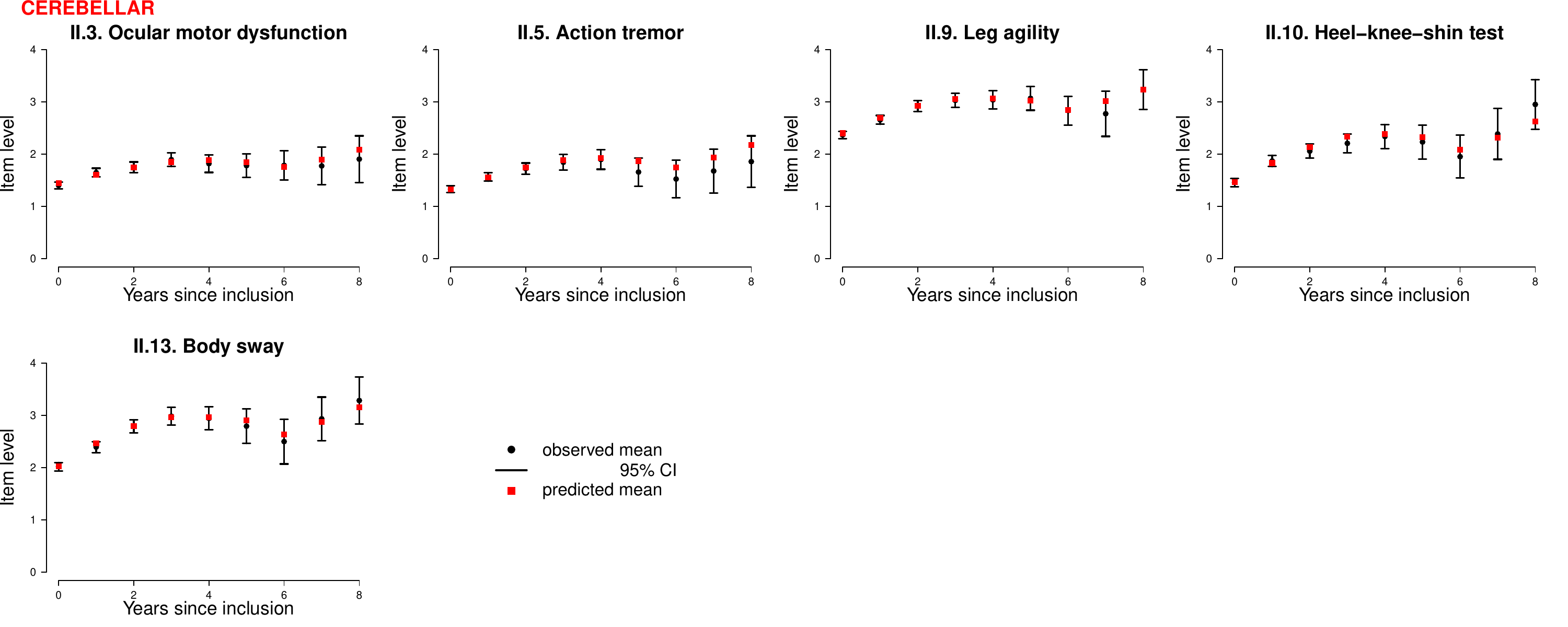}
    \includegraphics[width = 14cm, trim= 0cm 0cm 0cm 0cm, clip]{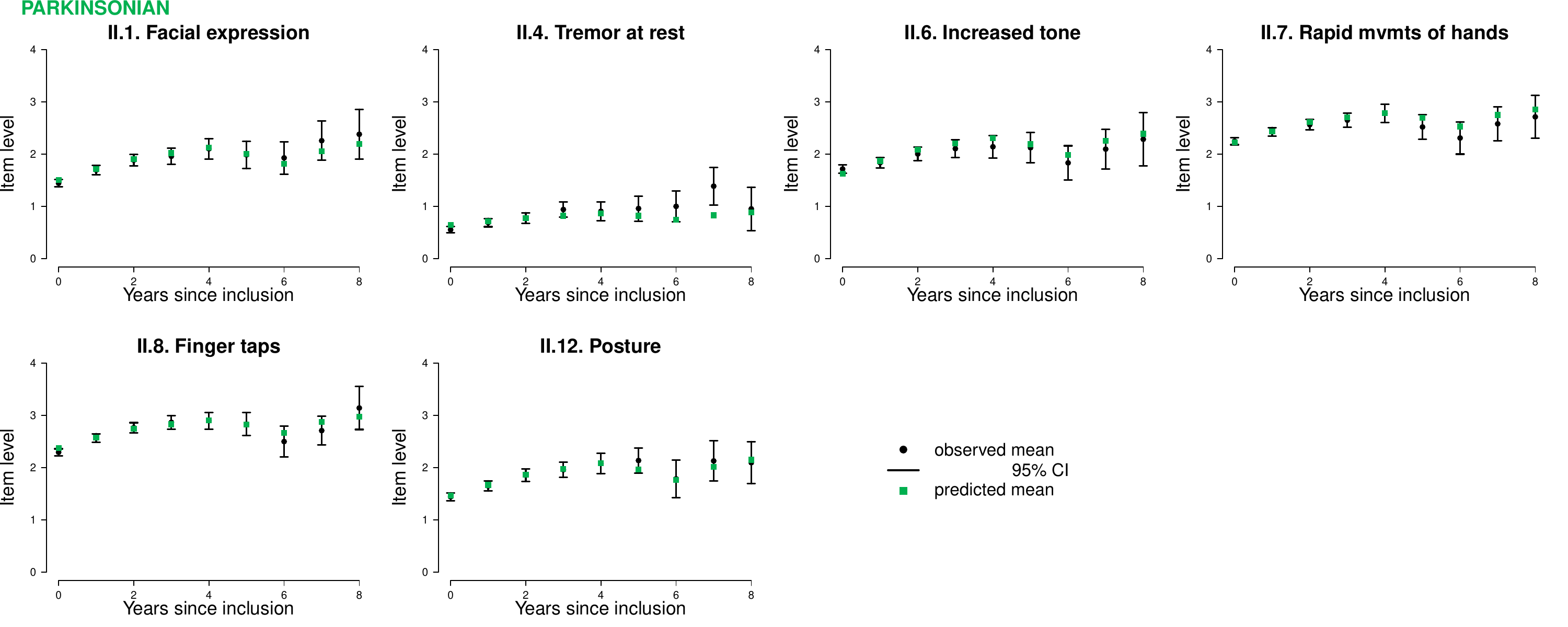}
    \label{fig:msa_fit_items}
\end{figure}

\begin{figure}[h!]
    \caption{\RVW{\textbf{Diagnostic plots assessing the fit of the survival part in the Step 2 estimated JLPMs.} Graphs compare the observed cumulative incidence curves and the mean predicted curves for both competing events — death and dropout.}}
    \vspace{.5cm}
    \centering
    \includegraphics[width = 11cm, trim= 0cm 0cm 0cm 0cm, clip]{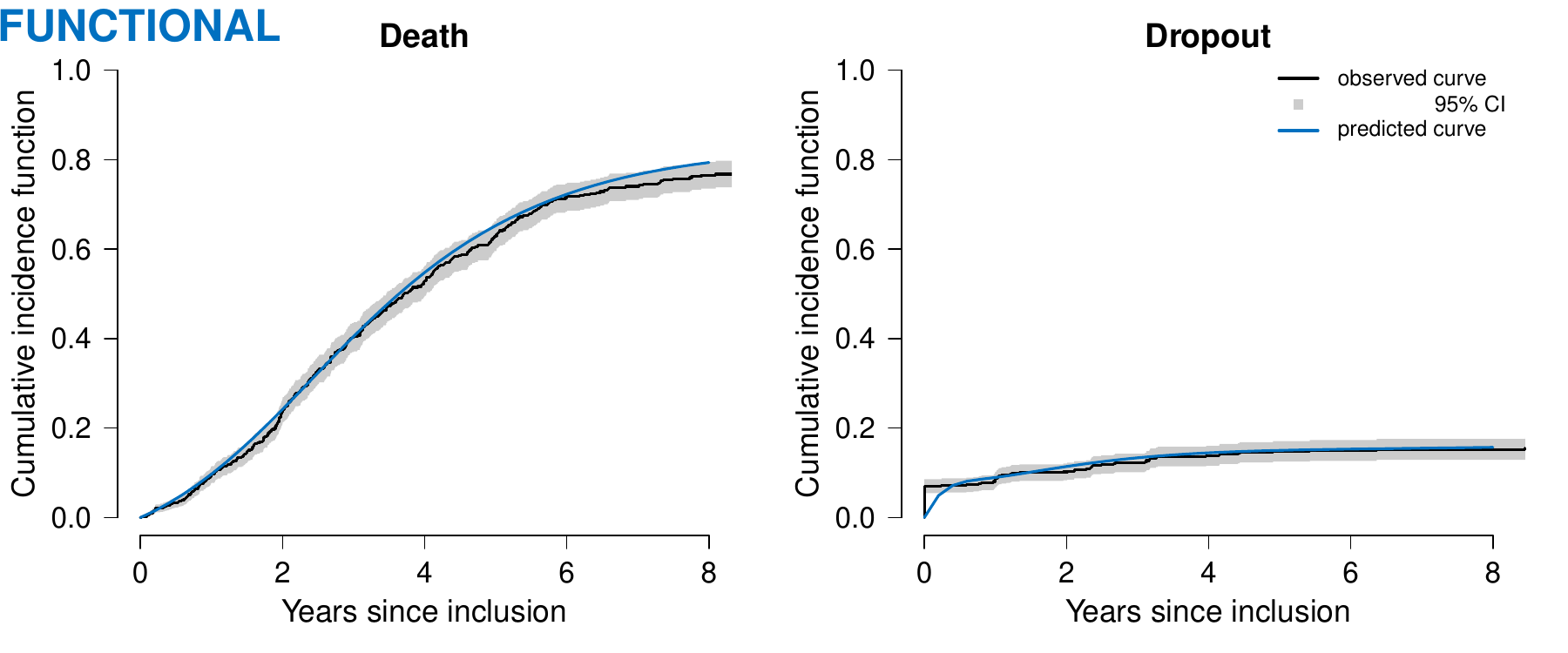}
    \includegraphics[width = 11cm, trim= 0cm 0cm 0cm 0cm, clip]{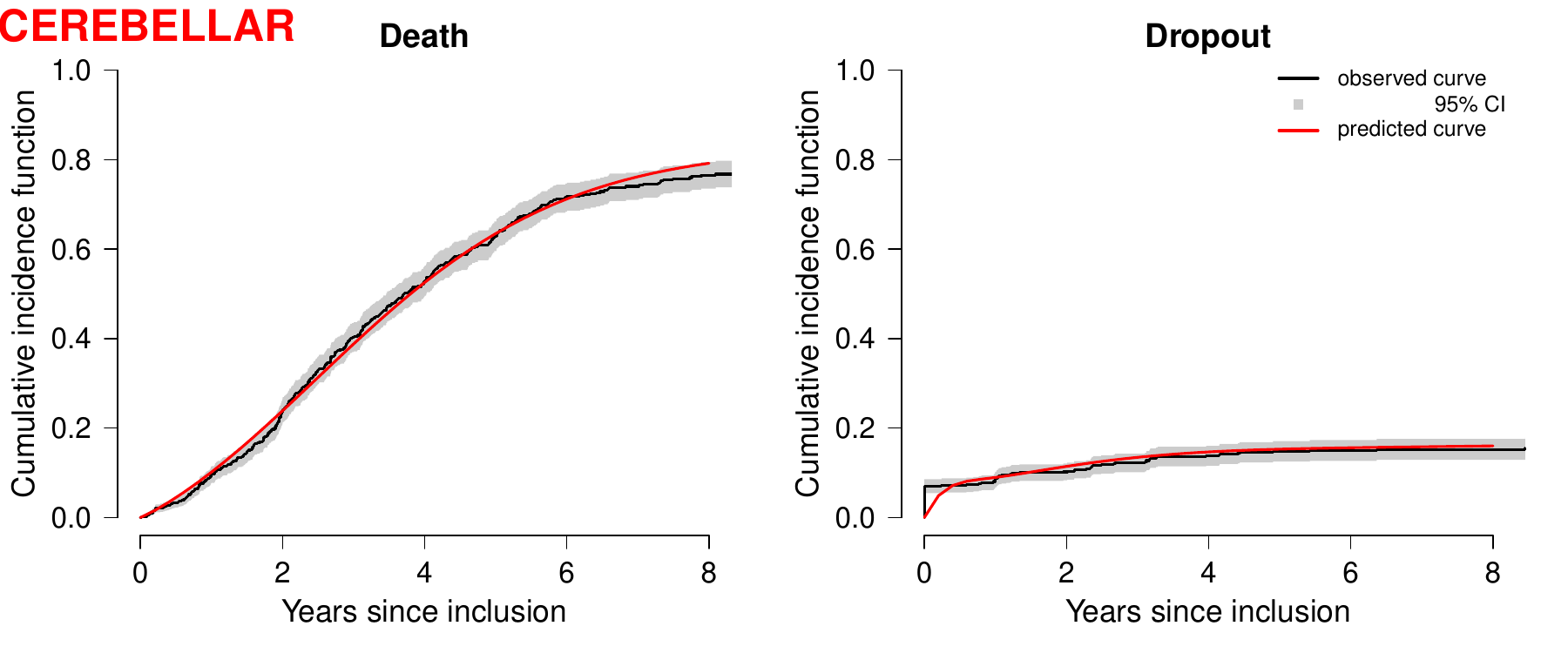}
    \includegraphics[width = 11cm, trim= 0cm 0cm 0cm 0cm, clip]{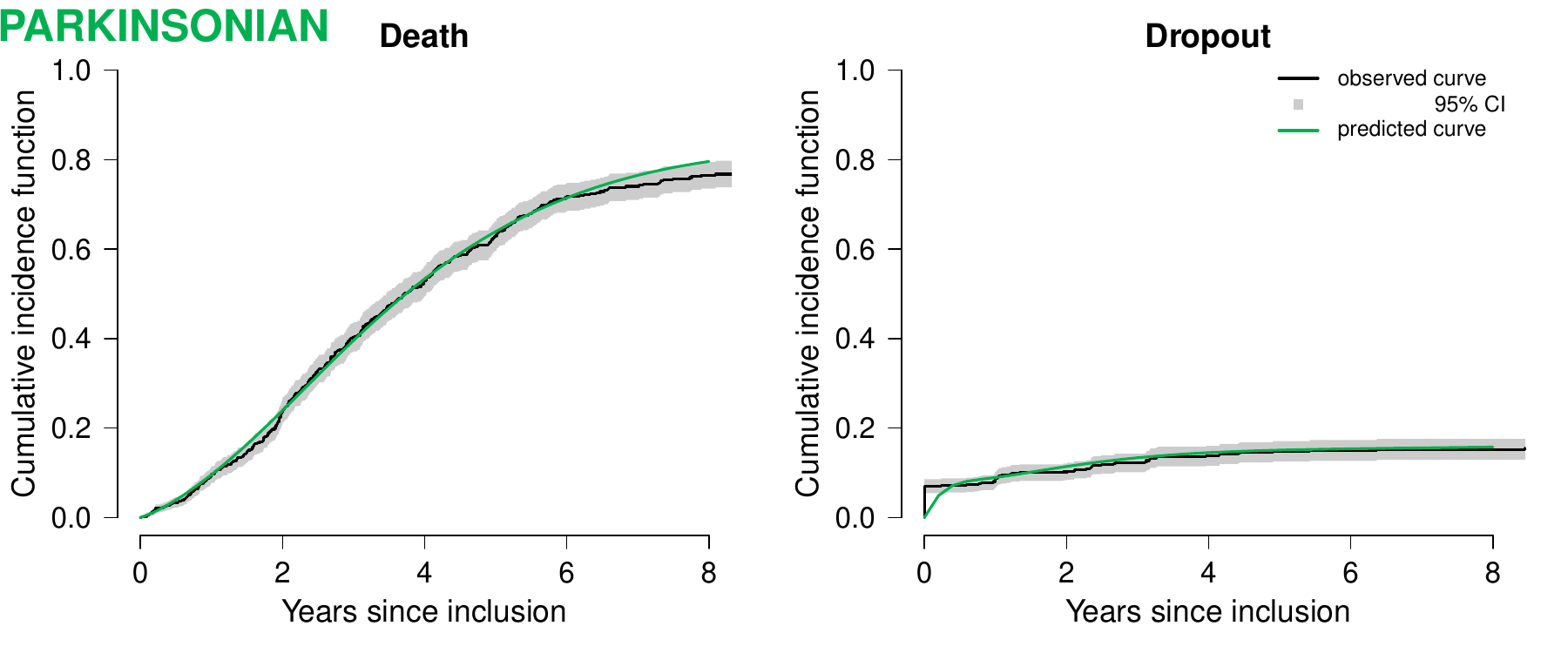}
    \label{fig:msa_fit_events}
\end{figure}

\RVW{Table \ref{tab:estimMSA:S2} reports the estimates of the JLPMs for the three dimensions. \\
Figure \ref{fig:msa_spider} displays the spider plots of item impairments over time according to covariate profiles.\\
Table \ref{tab:specifMSA:S2} reports the fit performances and point estimates of different JLPM specifications. \\
Figures \ref{fig:msa_fit_items} and \ref{fig:msa_fit_events} display diagnostic plots of the estimated JLPMs, assessing the fit of the longitudinal and survival parts, respectively.} \\

\noindent \RVW{Model fit was assessed for each subdimension through two visual comparisons between observed data and model-derived predictions on the analytical sample. 
Subject-specific random effects were predicted as the mode of the posterior distribution of the random effects given the observations. We computed individual predictions as the expected values conditional on these predicted subject-specific random effects.
\begin{itemize}
    \item For the longitudinal submodel, we compared the mean individual item predictions to the mean observations within thin time-intervals. \\ 
    We show the results in Figure \ref{fig:msa_fit_items}. The predicted means were close to the observed ones across the three subdimensions, supporting the adequacy of the estimated models to the observed repeated item data. 
    \item For the survival submodel, we compared the mean of the individual predicted cumulative incidence curves to the curves based on the non-parametric Aalen-Johansen estimator for both competing events — death and dropout. \\ 
    Figure \ref{fig:msa_fit_events} reports them for Step 2. The predicted curves aligned well with the observed ones across the three subdimensions, illustrating a satisfactory fit of the estimated models to the time-to-event data.
\end{itemize}
We note that the predictions were computed using the same data as employed for model estimation.}

\begin{table}[h!]
    \caption{\textbf{Estimates of the JLPMs for the MSA application in Step 2 - Sequencing} to describe functional, cerebellar, and parkinsonian subdimension sequences. For a sake of readability, parameters from the measurement submodels and random-effect parameters from the structural submodels are not listed.}
    \label{tab:estimMSA:S2}
    \vspace{.25cm}
    \centering
    \begin{sideways} 
    \small
    \begin{tabular}{lccc|ccc|ccc}
        \hline
        SUBDIMENSIONS $\rightarrow$ & \multicolumn{3}{c}{\cellcolor{bleu!60} FUNCTIONAL} & \multicolumn{3}{c}{\cellcolor{rouge!60} CEREBELLAR} & \multicolumn{3}{c}{\cellcolor{vert!60} PARKINSONIAN} \\
        PARAMETERS $\downarrow$ & estimate & se & p-value & estimate & se & p-value & estimate & se & p-value \\
        
        \multicolumn{10}{l}{\bf survival submodel} \\
            \multicolumn{10}{l}{\it  \hspace{.2cm} death event} \\
                \hspace{.5cm} baseline risk Weibull scale & 0.213 & 0.019 & $\mathbf{<0.001}$ & 0.296 & 0.022 & $\mathbf{<0.001}$ & 0.335 & 0.017 & $\mathbf{<0.001}$ \\
                \hspace{.5cm} baseline risk Weibull shape & 1.043 & 0.028 & $\mathbf{<0.001}$ & 1.064 & 0.028 & $\mathbf{<0.001}$ & 1.110 & 0.026 & $\mathbf{<0.001}$ \\
                \hspace{.5cm} current level association & 0.612 & 0.044 & $\mathbf{<0.001}$ & 0.396 & 0.041 & $\mathbf{<0.001}$ & 0.613 & 0.050 & $\mathbf{<0.001}$ \\
            \multicolumn{10}{l}{\it \hspace{.2cm} dropout event} \\
                \hspace{.5cm} baseline risk splines1 & 0.308 & 0.025 & $\mathbf{<0.001}$ & 0.291 & 0.022 & $\mathbf{<0.001}$ & 0.292 & 0.020 & $\mathbf{<0.001}$ \\
                \hspace{.5cm} baseline risk splines2 & 0.000 & 0.022 & 0.999 & 0.000 & 0.061 & 0.999 & 0.000 & 0.020 & 0.996 \\
                \hspace{.5cm} baseline risk splines3 & 0.207 & 0.036 & $\mathbf{<0.001}$ & 0.191 & 0.032 & $\mathbf{<0.001}$ & 0.197 & 0.032 & $\mathbf{<0.001}$ \\
                \hspace{.5cm} baseline risk splines4 & 0.397 & 0.076 & $\mathbf{<0.001}$ & 0.347 & 0.068 & $\mathbf{<0.001}$ & 0.377 & 0.063 & $\mathbf{<0.001}$ \\
                \hspace{.5cm} baseline risk splines5 & 0.145 & 0.321 & 0.651 & 0.123 & 0.310 & 0.692 & 0.143 & 0.295 & 0.629 \\
                \hspace{.5cm} baseline risk splines6 & -0.001 & 0.092 & 0.992 & $\mathbf{0.001}$ & 0.073 & 0.990 & -0.002 & 0.064 & 0.980 \\
                \hspace{.5cm} baseline risk splines7 & 1.047 & 0.276 & $\mathbf{<0.001}$ & 0.883 & 0.243 & $\mathbf{<0.001}$ & 0.985 & 0.234 & $\mathbf{<0.001}$ \\
                \hspace{.5cm} current level association & -0.095 & 0.082 & 0.248 & 0.012 & 0.074 & 0.875 & -0.143 & 0.080 & 0.073 \\

        \multicolumn{10}{l}{\bf structural submodel} \\
            \multicolumn{10}{l}{\it  \hspace{.2cm} time function} \\
                \hspace{.5cm} splines1 & 1.990 & 0.086 & $\mathbf{<0.001}$ & 1.953 & 0.178 & $\mathbf{<0.001}$ & 1.479 & 0.123 & $\mathbf{<0.001}$ \\
                \hspace{.5cm} splines2 & 3.875 & 0.122 & $\mathbf{<0.001}$ & 3.899 & 0.228 & $\mathbf{<0.001}$ & 2.800 & 0.138 & $\mathbf{<0.001}$ \\
                \hspace{.5cm} splines3 & 3.905 & 0.137 & $\mathbf{<0.001}$ & 3.965 & 0.276 & $\mathbf{<0.001}$ & 2.837 & 0.159 & $\mathbf{<0.001}$ \\
            \multicolumn{10}{l}{\it  \hspace{.2cm} adjustment covariates} \\
                \hspace{.5cm} year-delay since first symptom onset & 0.212 & 0.008 & $\mathbf{<0.001}$ & 0.157 & 0.018 & $\mathbf{<0.001}$ & 0.140 & 0.014 & $\mathbf{<0.001}$ \\
                \hspace{.5cm} sex, female & 0.452 & 0.060 & $\mathbf{<0.001}$ & 0.397 & 0.087 & $\mathbf{<0.001}$ & 0.282 & 0.072 & $\mathbf{<0.001}$ \\
                \hspace{.5cm} diagnosis subtype, MSA-C & 0.138 & 0.064 & $\mathbf{0.033}$ & 0.622 & 0.096 & $\mathbf{<0.001}$ & -1.307 & 0.092 & $\mathbf{<0.001}$ \\
                \hspace{.5cm} (sex, female):(diagnosis subtype, MSA-C) & -0.224 & 0.085 & $\mathbf{0.009}$ & - & - & - & - & - & - \\
                \hspace{.5cm} diagnosis certainty, possible & -0.305 & 0.055 & $\mathbf{<0.001}$ & -0.655 & 0.101 & $\mathbf{<0.001}$ & -0.563 & 0.087 & $\mathbf{<0.001}$ \\
                \hspace{.5cm} age at entry, 10 years gap since 65 & 0.202 & 0.027 & $\mathbf{<0.001}$ & 0.135 & 0.059 & $\mathbf{0.023}$ & 0.092 & 0.051 & 0.074 \\
                \hspace{.5cm} first symptom nature, dysautonomic & -0.746 & 0.101 & $\mathbf{<0.001}$ & -0.272 & 0.144 & 0.059 & -0.530 & 0.116 & $\mathbf{<0.001}$ \\
                \hspace{.5cm} first symptom nature, both & -0.297 & 0.049 & $\mathbf{<0.001}$ & -0.472 & 0.115 & $\mathbf{<0.001}$ & -0.290 & 0.087 & $\mathbf{<0.001}$ \\
                \hspace{.5cm} hypotension as first symtpom & 0.179 & 0.057 & $\mathbf{0.002}$ & -0.041 & 0.150 & 0.783 & -0.252 & 0.105 & $\mathbf{0.017}$ \\

        \hline
    \end{tabular}
    \end{sideways}
\end{table}

\newpage

\newpage

\begin{table}[h!]
    \caption{\RVW{\textbf{Fit performance (log-likelihood, AIC) and point estimate comparison between candidate JLPMs for Step 2.} For each subdimension, models differ by the association structure between the longitudinal and survival parts (no association for both events (1), or association through the dimension current level for death only (2) or both events (3)). Only association and fixed-effect parameters are listed.}}
    \label{tab:specifMSA:S2}
    \centering
    \begin{sideways} 
    {\fontsize{8.45pt}{11pt}\selectfont 
    \begin{tabular}{lccc|ccc|ccc}
        \hline
        \bf SUBDIMENSIONS $\rightarrow$ & \multicolumn{3}{c}{\cellcolor{bleu!60} FUNCTIONAL} & \multicolumn{3}{c}{\cellcolor{rouge!60} CEREBELLAR} & \multicolumn{3}{c}{\cellcolor{vert!60} PARKINSONIAN} \\
        \bf MODELS $\rightarrow$ & 1 & 2 & 3 & 1 & 2 & 3 & 1 & 2 & 3  \\
        number of parameters & 98 & 99 & 100 & 52 & 53 & 54 & 57 & 58 & 59 \\
        log-likelihood & -32269.85 & -32128.80 & -32128.11 & -14705.02 & -14636.43 & -14636.42 & -15850.94 & -15737.48 & -15735.85 \\
        Akaike Information Criterion (AIC) & 64735.70 & 64455.60 & 64456.22 & 29514.05 & 29378.86 & 29380.84 & 31815.87 & 31590.96 & 31589.70 \\
        \bf PARAMETERS $\downarrow$ & & & & & & & & & \\
        
        \multicolumn{10}{l}{\bf survival submodel} \\
            \multicolumn{10}{l}{\it  \hspace{.005cm} death event} \\
                \hspace{.1cm} current level association & - & 0.612 & 0.612 & - & 0.396 & 0.396 & - & 0.613 & 0.613 \\
            \multicolumn{10}{l}{\it \hspace{.005cm} dropout event} \\
                \hspace{.1cm} current level association & - & - & -0.095 & - & - & 0.011 & - & - & -0.143 \\

        \multicolumn{10}{l}{\bf structural submodel} \\
            \multicolumn{10}{l}{\it  \hspace{.005cm} time function} \\
                \hspace{.1cm} splines1 & 1.867 & 1.990 & 1.990 & 1.728 & 1.953 & 1.953 & 1.347 & 1.479 & 1.479 \\
                \hspace{.1cm} splines2 & 3.668 & 3.875 & 3.875 & 3.465 & 3.899 & 3.899 & 2.476 & 2.799 & 2.800 \\
                \hspace{.1cm} splines3 & 3.651 & 3.907 & 3.905 & 3.355 & 3.965 & 3.965 & 2.378 & 2.838 & 2.837 \\
            \multicolumn{10}{l}{\it  \hspace{.005cm} adjustment covariates} \\
                \hspace{.1cm} year-delay since first symptom onset & 0.198 & 0.212 & 0.212 & 0.146 & 0.157 & 0.157 & 0.131 & 0.140 & 0.140 \\
                \hspace{.1cm} sex, female & 0.485 & 0.452 & 0.452 & 0.461 & 0.397 & 0.397 & 0.245 & 0.282 & 0.282 \\
                \hspace{.1cm} diagnosis subtype, MSA-C & 0.149 & 0.138 & 0.138 & 0.607 & 0.622 & 0.622 & -1.306 & -1.308 & -1.307 \\
                \hspace{.1cm} (sex, female):(diagnosis subtype, MSA-C) & -0.491 & -0.225 & -0.224 & - & - & - & - & - & - \\
                \hspace{.1cm} diagnosis certainty, possible & -0.515 & -0.305 & -0.305 & -0.770 & -0.655 & -0.655 & -0.600 & -0.564 & -0.563 \\
                \hspace{.1cm} age at entry, 10 years gap since 65 & 0.257 & 0.203 & 0.202 & 0.078 & 0.135 & 0.135 & 0.166 & 0.091 & 0.092 \\
                \hspace{.1cm} first symptom nature, dysautonomic & -0.700 & -0.746 & -0.746 & -0.247 & -0.272 & -0.272 & -0.491 & -0.529 & -0.530 \\
                \hspace{.1cm} first symptom nature, both & -0.350 & -0.297 & -0.297 & -0.439 & -0.472 & -0.472 & -0.222 & -0.290 & -0.290 \\
                \hspace{.1cm} hypotension as first symtpom & 0.246 & 0.179 & 0.179 & -0.106 & -0.041 & -0.041 & -0.218 & -0.251 & -0.252 \\

        \hline
    \end{tabular}
    }
    \end{sideways}
\end{table}

\newpage

\newpage

\newpage

\subsubsection*{\RVW{4. Additional information for Step 3 - Staging}}

\begin{figure}[h!]
    \caption{\RVW{\textbf{Sequence of item impairments according to MSA stages for each subdimension with 95\% confidence intervals of the stage and item thresholds.} Each graph reports the item degradation according to the underlying subdimension continuum, along with the projection of the five MSA disease stage transitions, with $95\%$ confidence intervals obtained by parametric bootstrap over 1000 draws.}}
    \centering
    \includegraphics[width = 14cm, trim= 0cm 50cm 0cm .75cm, clip]{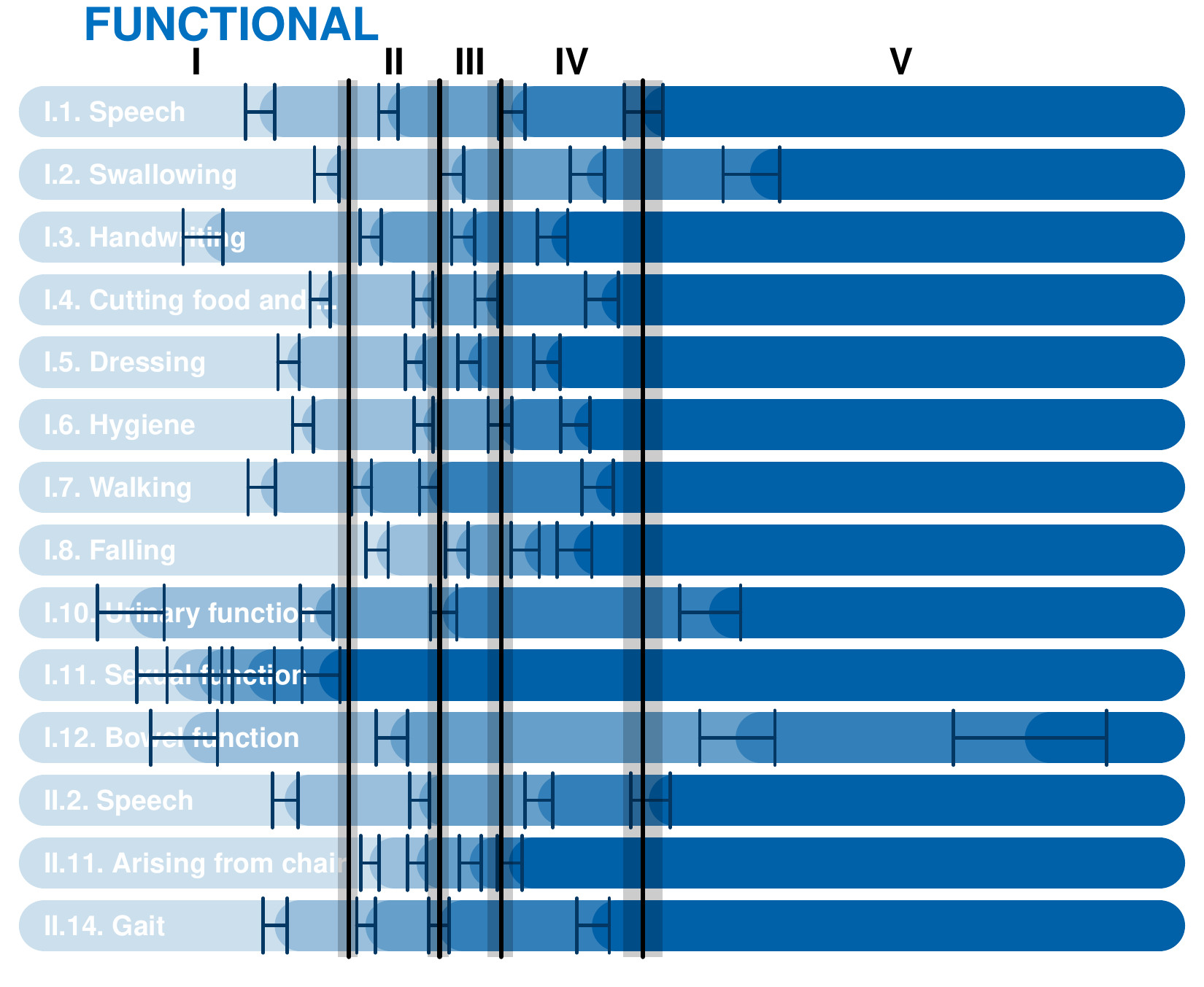}
    \includegraphics[width = 14cm, trim= 0cm 26cm 0cm 38.5cm, clip]{Graphs/IRT_UMSARS_Hierarchy_uncertainty.pdf}
    \includegraphics[width = 14cm, trim= 0cm 0cm 0cm 61cm, clip]{Graphs/IRT_UMSARS_Hierarchy_uncertainty.pdf}
    \label{fig:msa_hierarchy_uncertainty}
\end{figure}

\begin{figure}[h!]
    \caption{\RVW{\textbf{Diagnostic plots assessing the fit of the longitudinal part in the Step 3 estimated JLPMs.} Graphs compare the observed means of the dimension sum-scores or disease stages to the predicted means over a fine grid of times.}}
    \vspace{.5cm}
    \centering
    \includegraphics[width = 13cm, trim= 0cm 0cm 0cm 0cm, clip]{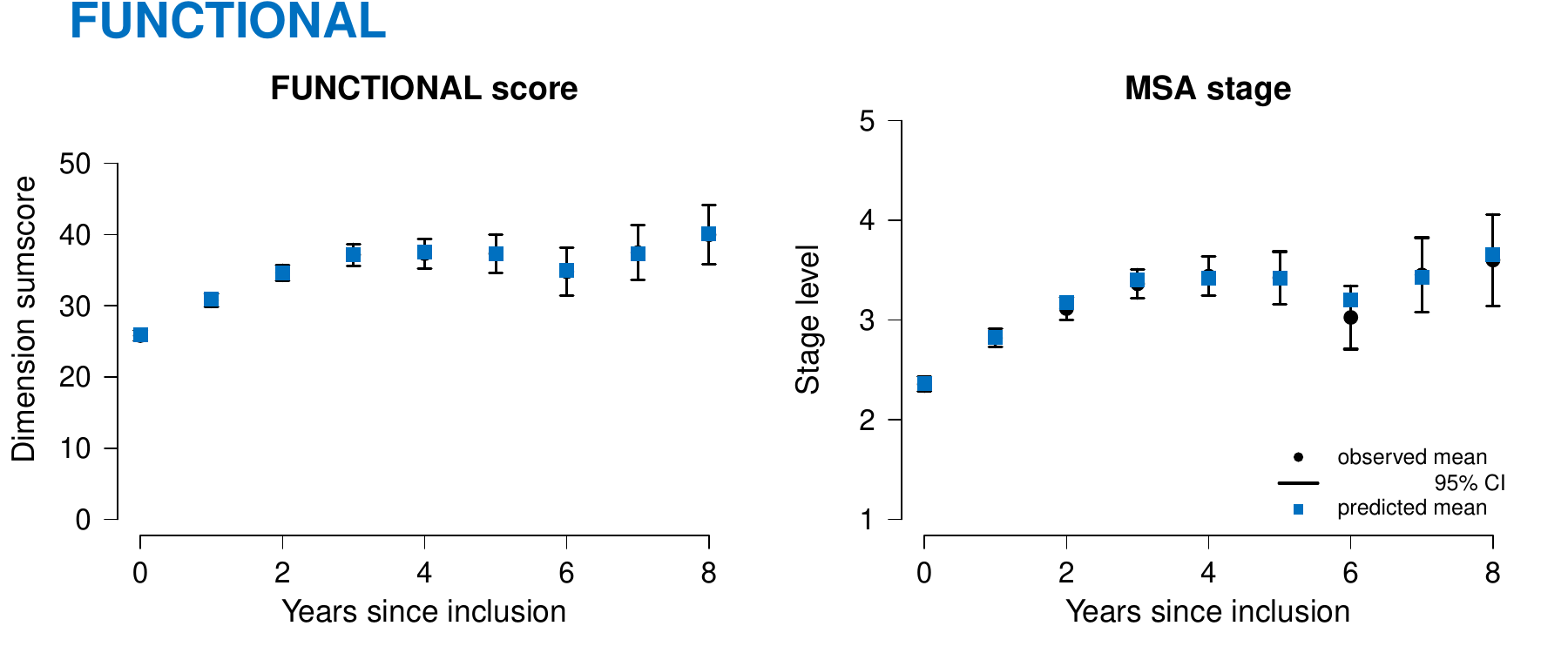}
    \includegraphics[width = 13cm, trim= 0cm 0cm 0cm 0cm, clip]{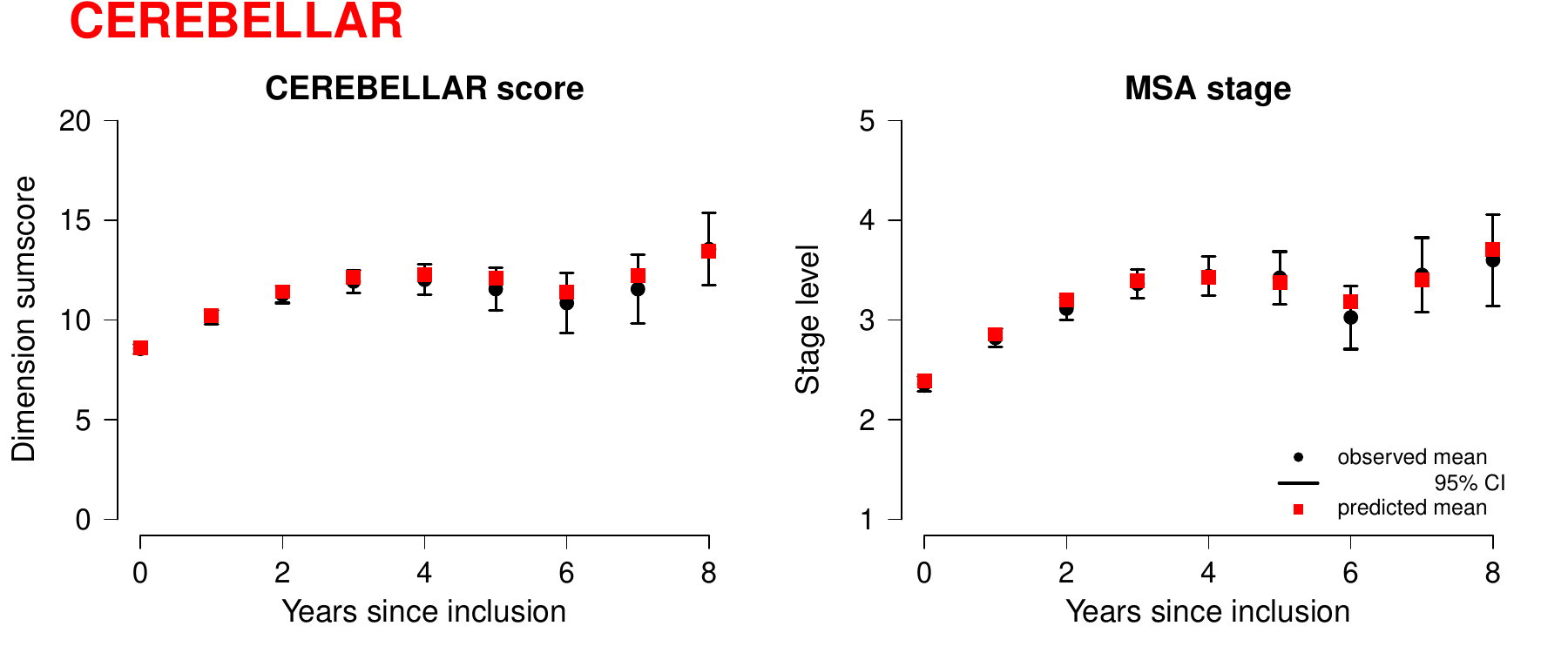}
    \includegraphics[width = 13cm, trim= 0cm 0cm 0cm 0cm, clip]{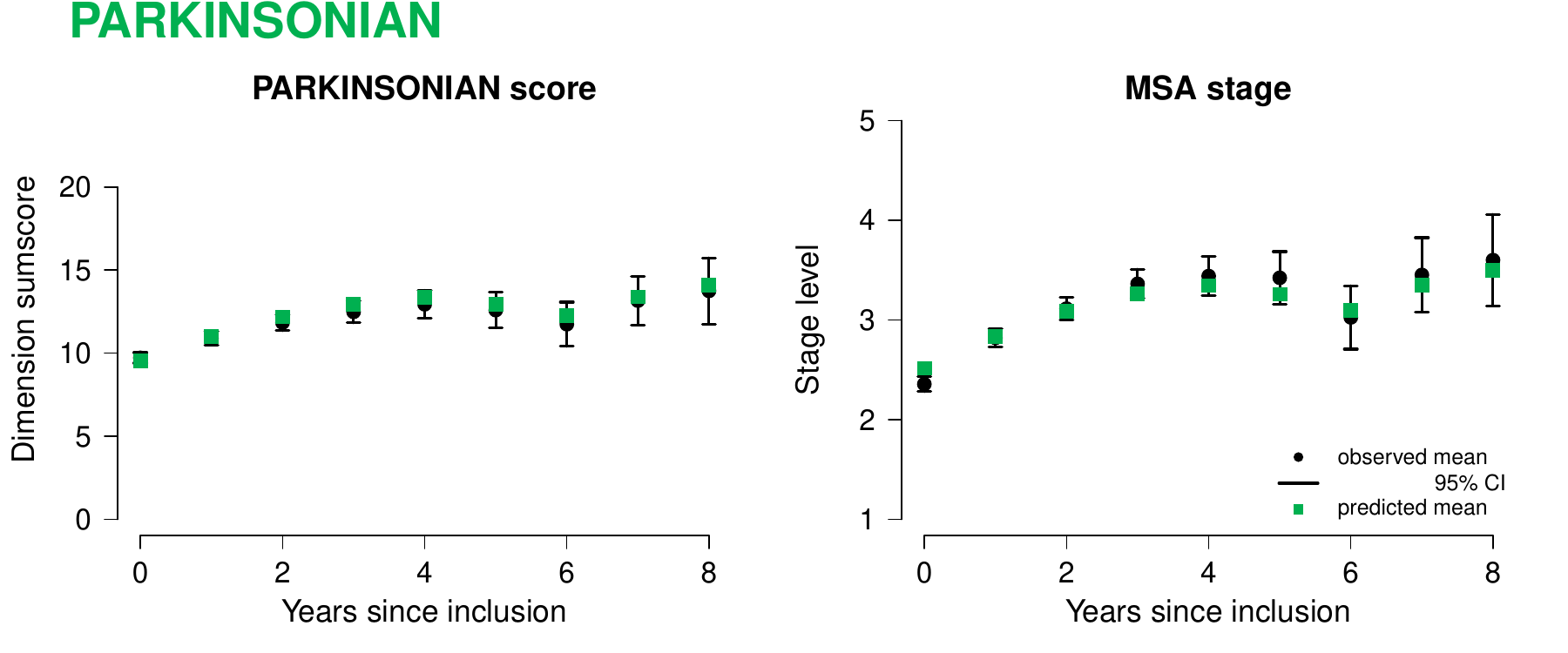}
    \label{fig:msa_fit_scorestage}
\end{figure}

\RVW{Figure \ref{fig:msa_hierarchy_uncertainty} reports the sequence of impairments with the disease stage anchoring with 95\% confidence intervals computed by parametric bootstrap over 1000 draws.\\
Figure \ref{fig:msa_fit_scorestage} displays the diagnostic plots assessing the fit of the longitudinal part in the estimated JLPMs.} \\

\noindent \RVW{In Step 3, model fit was assessed for each subdimension using a strategy similar to the one employed in Step 2. Specifically, we compared the mean individual marker predictions (i.e., dimension scores or disease stages) to the mean observations over a fine grid of times. \\ 
As shown in Figure \ref{fig:msa_fit_scorestage}, the predicted means were close to the observed ones across the three subdimensions, supporting the adequacy of the estimated models to the repeated marker data.}\\

\newpage

\newpage

\subsubsection*{\RVW{5. Additional information for Step 4 - Selecting}}

\RVW{The item rankings are reported in Table \ref{tab:fisher}, and in Table \ref{tab:fisher_uncertainty} with 95\% confidence intervals computed by parametric bootstrap over 1000 draws.}

\newpage

\begin{table}[h!]
    \caption{\textbf{Ranking of UMSARS-I and -II items per subdimension for the five MSA disease stages according to the item-specific Fisher information carried.} \textit{Info \%} is the percentage of Fisher information carried by the item, and \textit{Cum Info \%} is the cumulative percentage of Fisher information carried by the item and the most informative ones. \RVW{Cell coloring facilitates tracking changes in item rankings from one stage to another.}}
    \label{tab:fisher}
    \vspace{.1cm}
    \centering
    \begin{sideways} 
    \footnotesize
    \begin{tabular}{|c|l|c|c|c|c|c|c|c|c|c|c|c|c|c|c|}
        \hline
         & \multicolumn{3}{c|}{Stage I} & \multicolumn{3}{c|}{Stage II} & \multicolumn{3}{c|}{Stage III} & \multicolumn{3}{c|}{Stage IV} & \multicolumn{3}{c|}{Stage V} \\ \hline
        \multirow{3}{*}{\rotatebox[origin=c]{90}{Rank}} & \multirow{3}{*}{Item} & & Cum & & & Cum & & & Cum & & & Cum & & & Cum \\
          & & Info & Info & Item & Info & Info & Item & Info & Info & Item & Info & Info & Item & Info & Info \\
          & & \% & \% & & \% & \% & & \% & \% & & \% & \% & & \% & \% \\ \hline
        
        \multicolumn{16}{|l|}{\cellcolor{bleu!60}FUNCTIONAL subdimension} \\ \hline
        \cellcolor{bleu!60}1 & \cellcolor{colitem7!50}I.7 Walking & 14.8 & 14.8 & \cellcolor{colitem5!50}I.5 & 13.3 & 13.3 & \cellcolor{colitem5!50}I.5 & 16.3 & 16.3 & \cellcolor{colitem6!50}I.6 & 17.1 & 17.1 & \cellcolor{colitem2!50}I.2 & 27.0 & 27.0 \\ \hline
        \cellcolor{bleu!60}2 & \cellcolor{colitem5!50}I.5 Dressing & 14.7 & 29.5 & \cellcolor{colitem6!50}I.6 & 13.1 & 26.4 & \cellcolor{colitem6!50}I.6 & 16.3 & 32.6 & \cellcolor{colitem4!50}I.4 & 13.9 & 31.0 & \cellcolor{colitem11!50}I.12 & 18.9 & 45.9 \\ \hline
        \cellcolor{bleu!60}3 & \cellcolor{colitem6!50}I.6 Hygiene & 13.2 & 42.7 & \cellcolor{colitem7!50}I.7 & 12.9 & 39.3 & \cellcolor{colitem4!50}I.4 & 13.4 & 46.0 & \cellcolor{colitem5!50}I.5 & 13.1 & 44.1 & \cellcolor{colitem12!50}II.2 & 16.1 & 62.0 \\ \hline
        \cellcolor{bleu!60}4 & \cellcolor{colitem14!50}II.14 Gait & 12.8 & 55.5 & \cellcolor{colitem4!50}I.4 & 12.7 & 52.0 & \cellcolor{colitem13!50}II.11 & 13.0 & 59.0 & \cellcolor{colitem7!50}I.7 & 10.5 & 54.6 & \cellcolor{colitem1!40}I.1 & 14.2 & 76.2 \\ \hline
        \cellcolor{bleu!60}5 & \cellcolor{colitem4!50}I.4 Cutting food and ... & 8.3 & 63.8 & \cellcolor{colitem13!50}II.11 & 12.5 & 64.5 & \cellcolor{colitem14!50}II.14 & 9.0 & 68.0 & \cellcolor{colitem14!50}II.14 & 10.4 & 65.0 & \cellcolor{colitem9!50}I.10 & 11.5 & 87.6 \\ \hline
        \cellcolor{bleu!60}6 & \cellcolor{colitem1!40}I.1 Speech & 8.0 & 71.8 & \cellcolor{colitem14!50}II.14 & 12.0 & 76.5 & \cellcolor{colitem7!50}I.7 & 8.5 & 76.5 & \cellcolor{colitem13!50}II.11 & 7.3 & 72.3 & \cellcolor{colitem4!50}I.4 & 3.4 & 91.0 \\ \hline
        \cellcolor{bleu!60}7 & \cellcolor{colitem3!50}I.3 Handwriting & 7.6 & 79.4 & \cellcolor{colitem1!40}I.1 & 5.6 & 82.1 & \cellcolor{colitem12!50}II.2 & 5.6 & 82.1 & \cellcolor{colitem12!50}II.2 & 7.1 & 79.4 & \cellcolor{colitem8!50}I.8 & 2.7 & 93.8 \\ \hline
        \cellcolor{bleu!60}8 & \cellcolor{colitem12!50}II.2 Speech & 6.4 & 85.8 & \cellcolor{colitem12!50}II.2 & 5.5 & 87.6 & \cellcolor{colitem1!40}I.1 & 5.5 & 87.6 & \cellcolor{colitem1!40}I.1 & 6.9 & 86.2 & \cellcolor{colitem7!50}I.7 & 2.5 & 96.3 \\ \hline
        \cellcolor{bleu!60}9 & \cellcolor{colitem9!50}I.10 Urinary function & 3.9 & 89.7 & \cellcolor{colitem3!50}I.3 & 4.0 & 91.6 & \cellcolor{colitem3!50}I.3 & 4.1 & 91.7 & \cellcolor{colitem3!50}I.3 & 4.1 & 90.4 & \cellcolor{colitem14!50}II.14 & 2.2 & 98.5 \\ \hline
        \cellcolor{bleu!60}10 & \cellcolor{colitem10!50}I.11 Sexual function & 3.4 & 93.1 & \cellcolor{colitem8!50}I.8 & 2.8 & 94.4 & \cellcolor{colitem8!50}I.8 & 3.0 & 94.7 & \cellcolor{colitem2!50}I.2 & 3.4 & 93.8 & \cellcolor{colitem3!50}I.3 & 1.2 & 99.7 \\ \hline
        \cellcolor{bleu!60}11 & \cellcolor{colitem2!50}I.2 Swallowing & 2.2 & 95.3 & \cellcolor{colitem2!50}I.2 & 2.7 & 97.1 & \cellcolor{colitem2!50}I.2 & 2.7 & 97.4 & \cellcolor{colitem8!50}I.8 & 3.4 & 97.2 & \cellcolor{colitem6!50}I.6 & 0.3 & 100.0 \\ \hline
        \cellcolor{bleu!60}12 & \cellcolor{colitem11!50}I.12 Bowel function & 2.0 & 97.3 & \cellcolor{colitem9!50}I.10 & 1.3 & 98.4 & \cellcolor{colitem9!50}I.10 & 1.3 & 98.6 & \cellcolor{colitem9!50}I.10 & 1.5 & 98.7 & \cellcolor{colitem5!50}I.5 & 0.0 & 100.0 \\ \hline
        \cellcolor{bleu!60}13 & \cellcolor{colitem13!50}II.11 Arising from chair & 1.6 & 98.9 & \cellcolor{colitem10!50}I.11 & 0.9 & 99.3 & \cellcolor{colitem10!50}I.11 & 0.7 & 99.3 & \cellcolor{colitem11!50}I.12 & 0.9 & 99.5 & \cellcolor{colitem10!50}I.11 & 0.0 & 100.0 \\ \hline
        \cellcolor{bleu!60}14 & \cellcolor{colitem8!50}I.8 Falling & 1.1 & 100.0 & \cellcolor{colitem11!50}I.12 & 0.7 & 100.0 & \cellcolor{colitem11!50}I.12 & 0.7 & 100.0 & \cellcolor{colitem10!50}I.11 & 0.5 & 100.0 & \cellcolor{colitem13!50}II.11 & 0.0 & 100.0 \\ \hline

        \multicolumn{16}{|l|}{\cellcolor{rouge!60}CEREBELLAR subdimension} \\ \hline
        \cellcolor{rouge!60}1 & \cellcolor{colitem8!50}II.9 Leg agility & 36.6 & 36.6 & \cellcolor{colitem14!50}II.13 & 45.9 & 45.9 & \cellcolor{colitem14!50}II.13 & 45.8 & 45.8 & \cellcolor{colitem14!50}II.13 & 36.8 & 36.8 & \cellcolor{colitem5!50}II.5 & 48.1 & 48.1 \\ \hline
        \cellcolor{rouge!60}2 & \cellcolor{colitem14!50}II.13 Body sway & 31.2 & 67.8 & \cellcolor{colitem8!50}II.9 & 21.6 & 67.5 & \cellcolor{colitem8!50}II.9 & 20.8 & 66.6 & \cellcolor{colitem11!50}II.10 & 23.3 & 60.2 & \cellcolor{colitem2!50}II.3 & 34.4 & 82.6 \\ \hline
        \cellcolor{rouge!60}3 & \cellcolor{colitem2!50}II.3 Ocular motor dysfunction & 11.7 & 79.5 & \cellcolor{colitem11!50}II.10 & 18.6 & 86.1 & \cellcolor{colitem11!50}II.10 & 19.3 & 86.0 & \cellcolor{colitem8!50}II.9 & 21.9 & 82.1 & \cellcolor{colitem11!50}II.10 & 14.3 & 96.9 \\ \hline
        \cellcolor{rouge!60}4 & \cellcolor{colitem11!50}II.10 Heel-knee-skin test & 10.6 & 90.0 & \cellcolor{colitem5!50}II.5 & 9.5 & 95.7 & \cellcolor{colitem5!50}II.5 & 9.6 & 95.6 & \cellcolor{colitem5!50}II.5 & 12.2 & 94.3 & \cellcolor{colitem8!50}II.9 & 3.0 & 99.8 \\ \hline
        \cellcolor{rouge!60}5 & \cellcolor{colitem5!50}II.5 Action tremor & 10.0 & 100.0 & \cellcolor{colitem2!50}II.3 & 4.3 & 100.0 & \cellcolor{colitem2!50}II.3 & 4.4 & 100.0 & \cellcolor{colitem2!50}II.3 & 5.7 & 100.0 & \cellcolor{colitem14!50}II.13 & 0.2 & 100.0 \\ \hline

        \multicolumn{16}{|l|}{\cellcolor{vert!60}PARKINSONIAN subdimension} \\ \hline
        \cellcolor{vert!60}1 & \cellcolor{colitem11!50}II.8 Finger taps & 37.2 & 37.2 & \cellcolor{colitem9!50}II.7 & 23.5 & 23.5 & \cellcolor{colitem7!50}II.6 & 22.8 & 22.8 & \cellcolor{colitem7!50}II.6 & 23.0 & 23.0 & \cellcolor{colitem3!50}II.1 & 30.3 & 30.3 \\ \hline
        \cellcolor{vert!60}2 & \cellcolor{colitem9!50}II.7 Rapid movements of hands & 35.5 & 72.7 & \cellcolor{colitem7!50}II.6 & 22.5 & 46.0 & \cellcolor{colitem9!50}II.7 & 22.8 & 45.6 & \cellcolor{colitem9!50}II.7 & 21.9 & 44.9 & \cellcolor{colitem5!50}II.4 & 20.2 & 50.5 \\ \hline
        \cellcolor{vert!60}3 & \cellcolor{colitem3!50}II.1 Facial expression & 9.9 & 82.7 & \cellcolor{colitem11!50}II.8 & 22.0 & 68.0 & \cellcolor{colitem11!50}II.8 & 20.6 & 66.2 & \cellcolor{colitem11!50}II.8 & 19.5 & 64.4 & \cellcolor{colitem13!50}II.12 & 20.1 & 50.6 \\ \hline
        \cellcolor{vert!60}4 & \cellcolor{colitem13!50}II.12 Posture & 9.3 & 92.0 & \cellcolor{colitem3!50}II.1 & 17.6 & 85.6 & \cellcolor{colitem3!50}II.1 & 18.3 & 84.6 & \cellcolor{colitem3!50}II.1 & 19.2 & 83.6 & \cellcolor{colitem7!50}II.6 & 19.8 & 90.3 \\ \hline
        \cellcolor{vert!60}5 & \cellcolor{colitem7!50}II.6 Increased tone & 7.2 & 99.2 & \cellcolor{colitem13!50}II.12 & 13.3 & 98.9 & \cellcolor{colitem13!50}II.12 & 14.2 & 98.8 & \cellcolor{colitem13!50}II.12 & 15.0 & 98.6 & \cellcolor{colitem9!50}II.7 & 5.9 & 96.2 \\ \hline
        \cellcolor{vert!60}6 & \cellcolor{colitem5!50}II.4 Tremor at rest & 0.8 & 100.0 & \cellcolor{colitem5!50}II.4 & 1.1 & 100.0 & \cellcolor{colitem5!50}II.4 & 1.2 & 100.0 & \cellcolor{colitem5!50}II.4 & 1.4 & 100.0 & \cellcolor{colitem11!50}II.8 & 3.8 & 100.0 \\ \hline

    \end{tabular}
    \end{sideways}
\end{table}

\newpage

\begin{table}[h!]
    \caption{\RVW{\textbf{Uncertainty around the contribution of UMSARS-I and -II items per subdimension for the five MSA disease stages according to the item-specific Fisher information carried.} The table reports the mean percentages of Fisher information carried by each item over each disease stage, along with the corresponding $95\%$ confidence intervals obtained by parametric bootstrap over 1000 draws. Cell coloring facilitates comparisons with Table \ref{tab:fisher}.}}
    \label{tab:fisher_uncertainty}
    \vspace{.25cm}
    \centering
    \begin{sideways} 
    \footnotesize
    \begin{tabular}{|l|c|c|c|c|c|}
        \hline
        STAGES $\rightarrow$ & \multirow{2}{*}{Stage I} & \multirow{2}{*}{Stage II} & \multirow{2}{*}{Stage III} & \multirow{2}{*}{Stage IV} & \multirow{2}{*}{Stage V} \\
        ITEMS $\downarrow$ & & & & & \\ \hline
        \multicolumn{6}{|l|}{\cellcolor{bleu!60}FUNCTIONAL subdimension} \\ \hline
        \cellcolor{colitem1!40}I.1 Speech                    & 8.0 [7.3-8.7]       & 5.6 [5.1-6.2]      & 5.5 [5.0-6.1]        & 6.8 [6.2-7.6]      & 14.2 [11.8-16.5] \\ \hline
        \cellcolor{colitem2!50}I.2 Swallowing                & 2.2 [2.0-2.4]       & 2.7 [2.4-3.0]      & 2.7 [2.4-3.1]        & 3.4 [3.0-3.8]      & 26.5 [23.2-30.3] \\ \hline
        \cellcolor{colitem3!50}I.3 Handwriting               & 7.6 [7.1-8.2]       & 4.0 [3.6-4.5]      & 4.1 [3.6-4.6]        & 4.1 [3.7-4.6]      & 1.2 [0.8-1.7] \\ \hline
        \cellcolor{colitem4!50}I.4 Cutting food and ...      & 8.2 [7.2-9.2]       & 12.6 [11.5-13.8]   & 13.4 [12.2-14.7]     & 13.8 [12.8-14.9]   & 3.7 [1.8-6.0] \\ \hline
        \cellcolor{colitem5!50}I.5 Dressing                  & 14.7 [13.6-15.9]    & 13.3 [12.1-14.5]   & 16.3 [14.9-17.8]     & 13.2 [12.3-14.3]   & 0.0 [0.0-0.1] \\ \hline
        \cellcolor{colitem6!50}I.6 Hygiene                   & 13.1 [11.9-14.2]    & 13.0 [12.0-14.2]   & 16.2 [14.9-17.6]     & 17.1 [15.9-18.4]   & 0.4 [0.1-0.9] \\ \hline
        \cellcolor{colitem7!50}I.7 Walking                   & 14.9 [13.8-16.1]    & 13.0 [11.7-14.4]   & 8.6 [7.8-9.4]        & 10.5 [9.6-11.5]    & 2.8 [1.3-4.6] \\ \hline
        \cellcolor{colitem8!50}I.8 Falling                   & 1.1 [0.9-1.2]       & 2.8 [2.5-3.2]      & 3.0 [2.6-3.4]        & 3.4 [3.0-3.8]      & 2.7 [2.0-3.4] \\ \hline
        \cellcolor{colitem9!50}I.10 Urinary function         & 4.0 [3.6-4.4]       & 1.3 [1.1-1.6]      & 1.3 [1.1-1.5]        & 1.5 [1.3-1.8]      & 11.3 [9.9-12.8] \\ \hline
        \cellcolor{colitem10!50}I.11 Sexual function         & 3.4 [3.0-3.8]       & 1.0 [0.7-1.2]      & 0.7 [0.6-0.9]        & 0.5 [0.4-0.5]      & 0.0 [0.0-0.0] \\ \hline
        \cellcolor{colitem11!50}I.12 Bowel function          & 2.1 [1.9-2.3]       & 0.7 [0.6-0.9]      & 0.7 [0.5-0.8]        & 0.9 [0.7-1.1]      & 18.6 [15.6-22.2] \\ \hline
        \cellcolor{colitem12!50}II.2 Speech                  & 6.4 [5.9-7.0]       & 5.5 [5.0-6.0]      & 5.5 [5.0-6.1]        & 7.1 [6.4-7.8]      & 16.1 [13.4-18.4] \\ \hline
        \cellcolor{colitem13!50}II.11 Arising from chair     & 1.6 [1.2-2.1]       & 12.5 [11.4-13.7]   & 13.0 [11.8-14.4]     & 7.3 [6.7-8.0]      & 0.0 [0.0-0.0] \\ \hline
        \cellcolor{colitem14!50}II.14 Gait                   & 12.9 [11.8-14.0]    & 12.0 [10.9-13.2]   & 9.0 [8.2-9.9]        & 10.3 [9.5-11.2]    & 2.4 [1.2-3.8] \\ \hline
        \multicolumn{6}{|l|}{\cellcolor{rouge!60}CEREBELLAR subdimension} \\ \hline
        \cellcolor{colitem2!50}II.3 Ocular motor dysfunction & 11.8 [10.5-13.2]    & 4.4 [3.5-5.3]      & 4.4 [3.6-5.4]        & 5.6 [4.6-6.9]      & 34.0 [30.1-38.1] \\ \hline
        \cellcolor{colitem5!50}II.5 Action tremor            & 10.0 [9.1-10.9]     & 9.5 [8.1-11.1]     & 9.6 [8.2-11.2]       & 12.0 [10.2-14.0]   & 47.5 [43.5-51.5] \\ \hline
        \cellcolor{colitem8!50}II.9 Leg agility              & 36.6 [33.6-39.9]    & 21.6 [19.1-24.4]   & 20.8 [18.3-23.4]     & 21.8 [19.7-24.2]   & 3.3 [1.6-5.4] \\ \hline
        \cellcolor{colitem11!50}II.10 Heel-knee-skin test    & 10.5 [9.2-11.8]     & 18.6 [16.2-21.0]   & 19.3 [16.9-21.9]     & 23.1 [20.7-25.8]   & 15.0 [10.6-19.2] \\ \hline
        \cellcolor{colitem14!50}II.13 Body sway              & 31.1 [27.7-34.4]    & 45.9 [41.9-50.2]   & 45.9 [41.9-50.2]     & 37.4 [33.7-41.2]   & 0.2 [0.0-0.7] \\ \hline
        \multicolumn{6}{|l|}{\cellcolor{vert!60}PARKINSONIAN subdimension} \\ \hline
        \cellcolor{colitem3!50}II.1 Facial expression        & 9.9 [8.4-11.5]      & 17.6 [15.8-19.6]   & 18.3 [16.5-20.3]     & 19.2 [17.3-21.2]   & 30.2 [26.9-33.6] \\ \hline
        \cellcolor{colitem5!50}II.4 Tremor at rest           & 0.8 [0.7-0.9]       & 1.1 [0.8-1.4]      & 1.2 [0.9-1.7]        & 1.4 [1.1-2.0]      & 19.9 [16.1-24.9] \\ \hline
        \cellcolor{colitem7!50}II.6 Increased tone           & 7.2 [5.6-8.8]       & 22.5 [20.3-24.7]   & 22.8 [20.8-25.0]     & 23.0 [21.0-25.1]   & 19.9 [16.5-23.3] \\ \hline
        \cellcolor{colitem9!50}II.7 Rapid movements of hands & 35.5 [33.3-38.1]    & 23.5 [21.4-25.9]   & 22.8 [20.8-25.1]     & 21.9 [20.0-24.1]   & 6.1 [4.1-8.4] \\ \hline
        \cellcolor{colitem11!50}II.8 Finger taps             & 37.3 [34.5-40.1]    & 22.0 [20.1-24.2]   & 20.6 [18.8-22.5]     & 19.5 [17.8-21.3]   & 3.9 [2.4-5.4] \\ \hline
        \cellcolor{colitem13!50}II.12 Posture                & 9.3 [8.1-10.5]      & 13.3 [11.9-14.9]   & 14.2 [12.7-15.9]     & 15.0 [13.4-16.7]   & 20.0 [17.4-22.4] \\ \hline
    \end{tabular}
    \end{sideways}
\end{table}
